\documentclass[aps,pre,twocolumn,showkeys,superscriptaddress]{revtex4}
\usepackage{epsfig,color}
\usepackage{times}
\usepackage{amsmath}

\begin{document}

\title{Cyclic dominance in evolutionary games: A review}

\author{Attila Szolnoki}
\email{szolnoki.attila@ttk.mta.hu}
\affiliation{Institute of Technical Physics and Materials Science, Research Centre for Natural Sciences, Hungarian Academy of Sciences, P.O. Box 49, H-1525 Budapest, Hungary}

\author{Mauro Mobilia}
\email{m.mobilia@leeds.ac.uk}
\affiliation{Department of Applied Mathematics, School of Mathematics, University of Leeds, Leeds LS2 9JT, U.K.}

\author{Luo-Luo Jiang}
\email{jiangluoluo@wzu.edu.cn}
\affiliation{College of Physics and Electronic Information Engineering, Wenzhou University, 325035 Wenzhou, China}

\author{Bartosz Szczesny}
\affiliation{Department of Applied Mathematics, School of Mathematics, University of Leeds, Leeds LS2 9JT, U.K.}

\author{Alastair M. Rucklidge}
\affiliation{Department of Applied Mathematics, School of Mathematics, University of Leeds, Leeds LS2 9JT, U.K.}

\author{Matja\v{z} Perc}
\email{matjaz.perc@uni-mb.si}
\affiliation{Faculty of Natural Sciences and Mathematics, University of Maribor, Koro\v{s}ka cesta 160, SI-2000 Maribor, Slovenia}

\begin{abstract}
Rock is wrapped by paper, paper is cut by scissors, and scissors are crushed by rock. This simple game is popular among children and adults to decide on trivial disputes that have no obvious winner, but cyclic dominance is also at the heart of predator-prey interactions, the mating strategy of side-blotched lizards, the overgrowth of marine sessile organisms, and the competition in microbial populations. Cyclical interactions also emerge spontaneously in evolutionary games entailing volunteering, reward, punishment, and in fact are common when the competing strategies are three or more regardless of the particularities of the game. Here we review recent advances on the rock-paper-scissors and related evolutionary games, focusing in particular on pattern formation, the impact of mobility, and the spontaneous emergence of cyclic dominance. We also review mean-field and zero-dimensional rock-paper-scissors models and the application of the complex Ginzburg-Landau equation, and we highlight the importance and usefulness of statistical physics for the successful study of large-scale ecological systems. Directions for future research, related for example to dynamical effects of coevolutionary rules and invasion reversals due to multi-point interactions, are outlined as well.
\end{abstract}

\keywords{cyclical interactions, pattern formation, phase transitions, mobility, coevolution, complex Ginzburg-Landau equation, self-organization}

\maketitle

\section{Introduction}
\label{intro}

Games of cyclic dominance play a prominent role in explaining the intriguing biodiversity in nature \cite{czaran_pnas02, kerr_n02, reichenbach_n07}, and they are also able to provide insights into Darwinian selection \cite{maynard_n73}, as well as into structural complexity \cite{watt_je47} and prebiotic evolution \cite{rasmussen_s04}. Cyclical interactions have been observed in different ecological systems. Examples include marine benthic systems \cite{jackson_pnas75}, plant systems \cite{taylor_am90, silvertown_je92, durrett_tpb98, lankau_s07, cameron_jecol09}, and microbial populations \cite{durrett_jtb97, kerr_n02,kirkup_n04,neumann_gf_bs10,nahum_pnas11}. Cyclic dominance also plays an important role in the mating strategy of side-blotched lizards \cite{sinervo_n96}, the overgrowth of marine sessile organisms \cite{burrows_mep98}, the genetic regulation in the repressilator \cite{elowitz_n00}, and in explaining the oscillating frequency of lemmings \cite{gilg_s03} and of the Pacific salmon \cite{guill_jtb11}. The list of examples where the puzzle of biological diversity can be explained by the interaction topology of the food web among those who struggle for life is indeed impressively long and inspiring \cite{berlow_jae04,stouffer_s12}.

In addition to the closed loops of dominance in the food webs that govern the evolution of competing species and strategies, experiments have shown that the key to the sustenance of biodiversity is often spatial structure. In vitro experiments with \textit{Escherichia coli}, for example, have shown that arranging the bacteria on a Petri dish is crucial for keeping all three competing strands alive \cite{kerr_n02, kerr_n06, weber_jrsi14}. On the other hand, in vivo experiments with bacterial colonies in the intestines of co-caged mice, which can be considered as locally well-mixed populations, have revealed that mobility alone is sufficient to maintain coexistence \cite{kirkup_n04}. Solely because the bacteria were able to migrate from one mice to another, the biodiversity was preserved. Motivated by such intriguing experimental observations, theoretical explorations of the relevance of spatial structure in evolutionary games have also received ample attention in the recent past \cite{szabo_pr07, roca_plr09, frey_pa10, perc_bs10, szollosi_pre08, perc_jrsi13}.

The importance of spatial structure has been put into the spotlight by the discovery that interaction networks like the square lattice can promote the evolution of cooperation in the prisoner's dilemma game through the mechanism that is now known as network reciprocity \cite{nowak_n92b, nowak_s06}. In essence, network reciprocity relies on the fact that cooperators do best if they are surrounded by other cooperators. More generally, the study of spatial games has revealed the need for realistic models to go beyond well-mixed populations \cite{mestertong_n10, bednarik_prsb14, van-doorn_jtb14}. To understand the fascinating diversity in nature and to reveal the many hidden mechanisms that sustain it, we cannot avoid the complexity of such systems. Cyclical interactions are characterized by the presence of strong fluctuations, unpredictable and nonlinear dynamics, multiple scales of space and time, and frequently also some form of emergent structure. Since many of these properties are inherent also to problems of nonequilibrium statistical physics, theoretical research on spatial rock-paper-scissors (RPS) and related games of cyclic dominance has a long and fruitful history \cite{tainaka_prl89, frachebourg_prl96, frachebourg_pre96, szabo_pre99, frean_prsb01, reichenbach_pre06, mobilia_pre06, szabo_jtb07, reichenbach_prl07, reichenbach_prl08, peltomaki_pre08, peltomaki_pre08b, berr_prl09, he_q_pre10, wang_wx_pre10b, ni_x_c10, mobilia_jtb10, wang_wx_pre11, mathiesen_prl11, avelino_pre12, jiang_ll_pla12, szolnoki_pre14b, roman_jsm12, avelino_pre12b, juul_pre12, roman_pre13, vukov_pre13, juul_pre13,maciejewski_pcbi14}, much of which is firmly rooted in methods of statistical physics that can inform relevantly on the outcome of evolutionary games on structured populations.

\begin{figure*}
\centerline{\epsfig{file=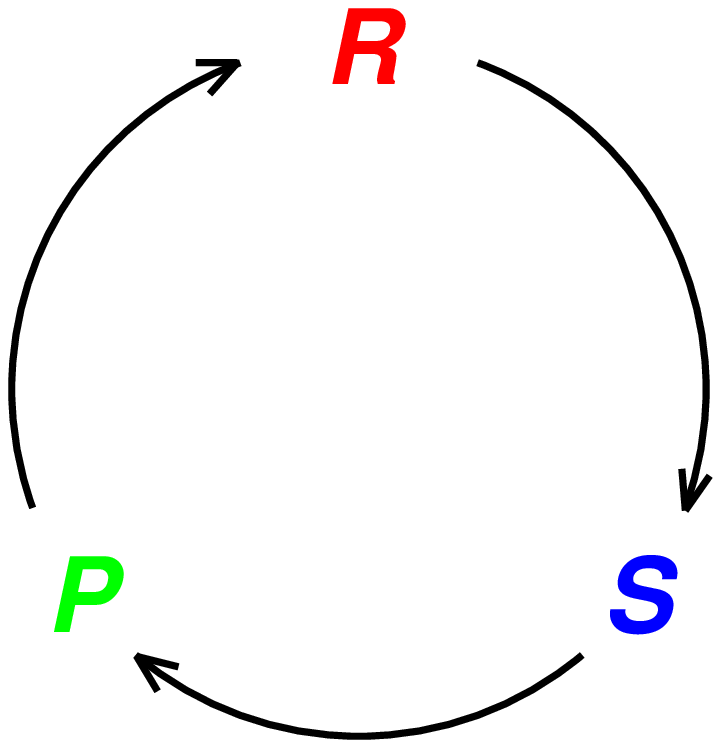,width=2.1cm}\epsfig{file=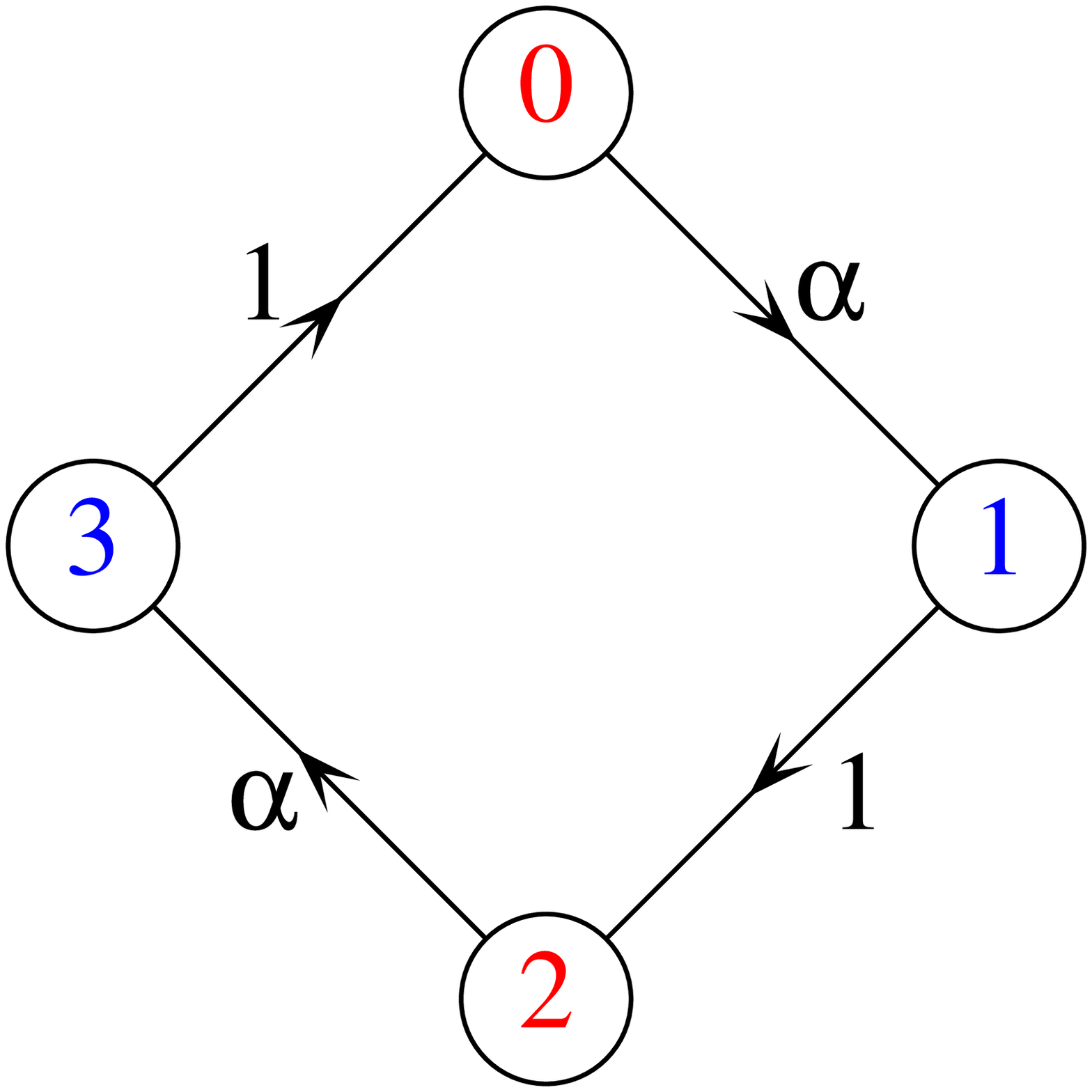,width=3.5cm}\epsfig{file=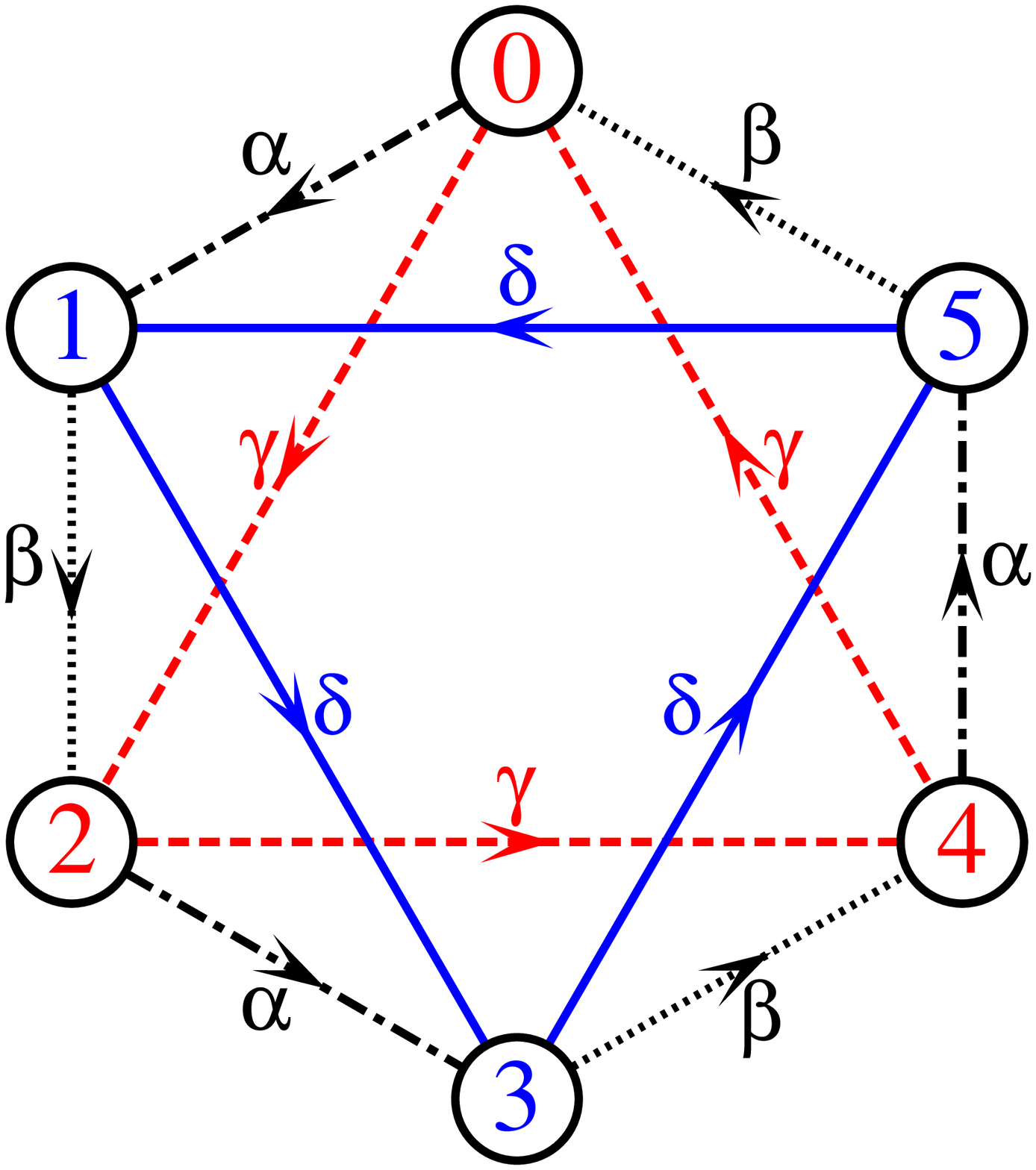,width=4.4cm}\epsfig{file=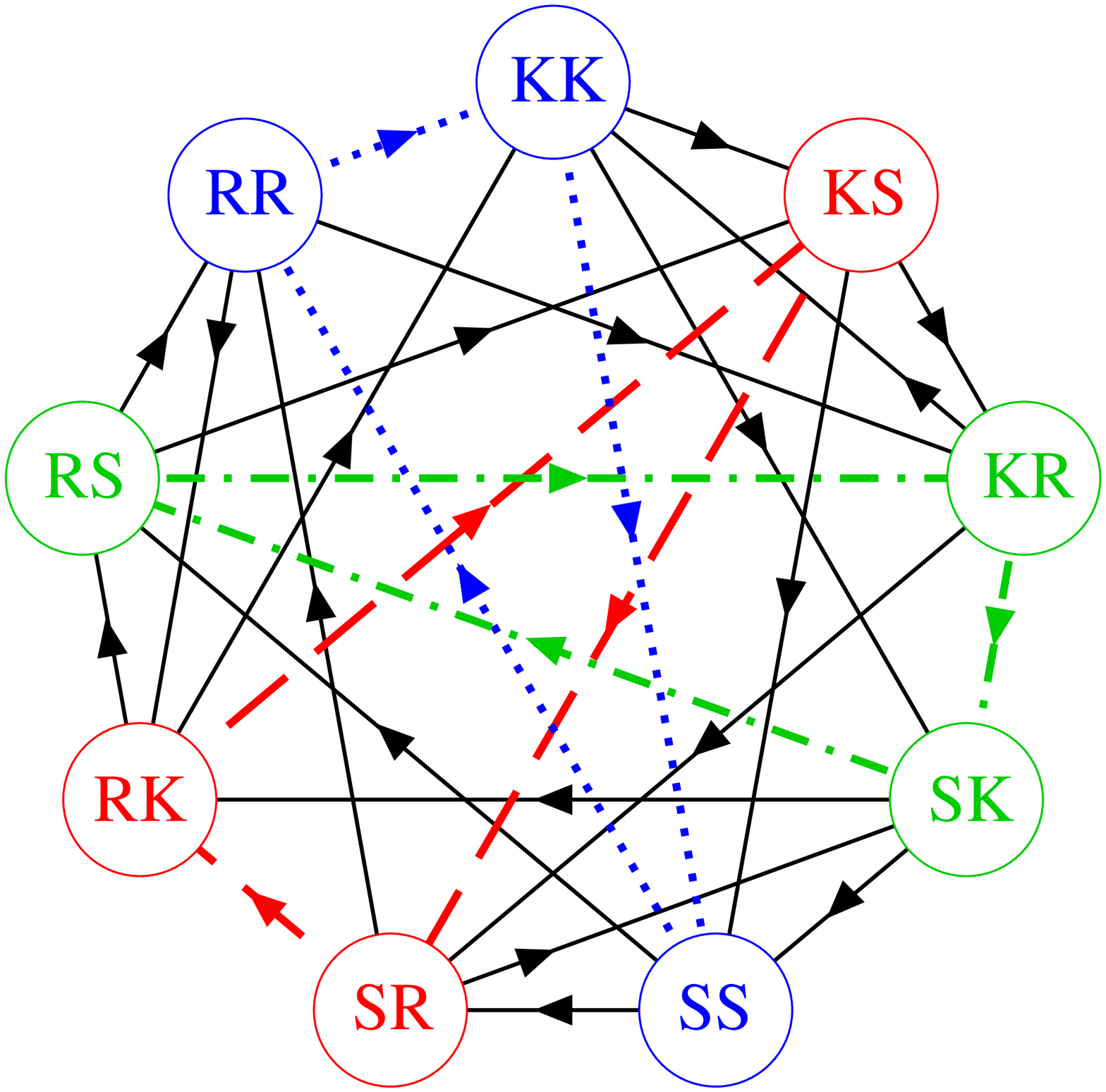,width=5.5cm}}
\caption{Examples of food webs of competing strategies. The number of competing strategies and the topological complexity of the food web increases from left to right, and so does the complexity and number of possible stable solutions: (a) The traditional RPS food web. Here rock (R) is wrapped by paper, paper (P) is cut by scissors, and scissors (S) are broken by a rock. (b) Four-strategy food web with different invasion rates between predator-prey pairs, which already supports the emergence of defensive alliances \cite{szabo_pre08}. (c) Six-species food web with heterogeneous invasion rates, supporting a multitude of different defensive alliances as well as noise-guided evolution \cite{perc_njp07b}. (d) Topologically complex food web describing bacterial warfare with two toxins. Here nine different strains compete, giving rise to a multitude of stable solutions that amaze with their complexity and beauty \cite{szabo_jtb07}. Three-strategy alliances that can successfully avert an external invasion are denoted by broken colored lines.}
\label{web}
\end{figure*}

Cyclical interactions are in many ways the culmination of evolutionary complexity, and they may also arise spontaneously in evolutionary games where the number of competing strategies is three or more. For example, cyclic dominance has been observed in public goods games with volunteering \cite{hauert_s02, semmann_n03}, peer punishment \cite{hauert_s02, helbing_ploscb10, amor_pre11, bednarik_prsb14}, pool punishment \cite{sigmund_n10, szolnoki_pre11} and reward \cite{hauert_jtb10, szolnoki_epl10, sigmund_pnas01}, but also in pairwise social dilemmas with coevolution \cite{szolnoki_epl09, szolnoki_pre10b} or with jokers \cite{requejo_pre12b}. Counterintuitive complex phenomena that are due to cyclic dominance include the survival of the weakest \cite{frean_prsb01, berr_prl09}, the emergence of labyrinthine clustering \cite{juul_pre13}, and the segregation along interfaces that have internal structure \cite{avelino_pre14}, to name but a few examples. Importantly, the complexity of solutions increases further and fast if the competing strategies are more than three. Figure~\ref{web} depicts a transition towards more and more complex food webs from left to right, which support the formation of defensive alliances \cite{szabo_jpa05} and subsystem solutions \cite{szabo_pre01a} that are not possible in the simplest RPS game.

Although the outcome of evolutionary games with cyclic dominance may depend on the properties of the interaction network, the topology of the food web, and on the number of competing strategies, there nevertheless exists fascinatingly robust universal behavior that is independent of model particularities. While we may have only begun to scratch the surface of the actual importance of cyclic dominance in nature \cite{kaspari_s14, lebrun_s14}, the time is ripe for a survey of theoretical advances that may guide future experimental efforts. Like by the evolution of cooperation \cite{axelrod_84}, in games of cyclic dominance too the evolutionary game theory \cite{maynard_82, hofbauer_98, mestertong_01, nowak_06, sigmund_10} is proving to be the theory of choice for the study of these beautifully simple yet fascinating systems.

The organization of this review is as follows. In Section~II we focus on mean-field and zero-dimensional RPS models to give a comprehensive prologue to the overview of the spatial RPS game in Section~III. Within the latter, we separately review the impact of different interaction networks and mobility, and we describe the derivation and the proper use of the complex Ginzburg-Landau equation to describe the spatiotemporal properties and the stability of the spiraling patterns. In Section~IV, we depart from the RPS game where the cyclic dominance is explicitly encoded in the food web by considering social dilemmas and other evolutionary games. We separately review games where cyclical interactions emerge spontaneously, the role of time-dependent interactions, the impact of voluntary participation, and we survey evolutionary games where an alliance between two strategies acts as an additional independent strategy. From there, we focus on cyclical interaction between four and more competing strategies, which opens the door towards reviewing the emergence of defensive alliances and large networks of competitive species. We conclude with a summary and an outlook towards promising future research efforts.

\section{Cyclic dominance in well-mixed systems}
\label{non-spatial}

While the emphasis of this review is on cyclic dominance in structured populations, it is useful to start by considering the simplest setting to study RPS games which is the mean field limit of well-mixed populations where all individuals can interact with each other.

\begin{figure}
\begin{center}
\includegraphics[scale=1]{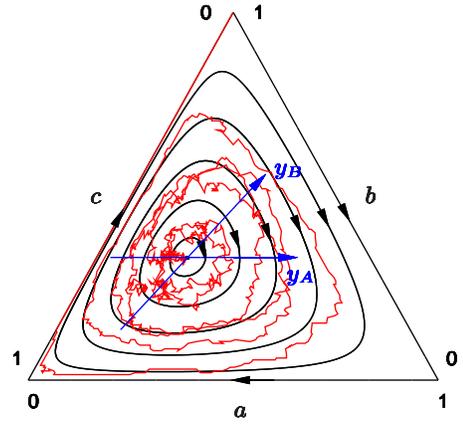}
\caption{Phase portrait of the RPS dynamics
with asymmetric dominance-replacement rates $z_A=2,~z_B=1,~z_C=1.5$ in a well-mixed population. The mean field rate equations predict closed orbits,
here shown in black solid lines. Their linearization around the neutrally stable
coexistence steady state ${\bf s}^*=(2/9,1/3,4/9)$ transform the orbits  into circles in the
 coordinates $y_A,y_B$ (blue). The red flow
that erratically spirals out from the coexistence fixed point
is a single trajectory obtained by simulating the stochastic dynamics in a finite population of
size $N=200$. This illustrates how demographic fluctuations are responsible
for the extinction of two species (here $A$ and $C$)
 after a time of order $N$. See the main text for details, as well as \cite{reichenbach_pre06}, from where this figure is adapted.}
\label{MF_portrait}
\end{center}
\end{figure}

A generic RPS game between individuals of three species, $A$, $B$ and $C$, in a population of total size $N$ entails cyclic competition and reproduction according to the following reaction schemes:
\begin{eqnarray}
\label{dom1}
AB &\stackrel{p}{\longrightarrow}& A\emptyset\,,\quad \,\,BC \stackrel{p}{\longrightarrow} B\emptyset\,,\quad \,CA \stackrel{p}{\longrightarrow} C\emptyset,\\
\label{dom2}
AB &\stackrel{z}{\longrightarrow}& AA\,,\quad BC \stackrel{z}{\longrightarrow} BB\,,\quad CA \stackrel{z}{\longrightarrow} CC,\\
\label{rep}
A\emptyset \,&\stackrel{q}{\longrightarrow}& AA\,,\quad B\emptyset \,\,\stackrel{q}{\longrightarrow} BB\,,\quad C\emptyset \,\stackrel{q}{\longrightarrow} CC,
\label{eq:reactions}
\end{eqnarray}
where the reactions (\ref{dom1}) and (\ref{dom2}) describe two forms of cyclic dominance of species $A$ over $B$, $B$ over $C$, and in turn $C$ over $A$ and (\ref{rep}) accounts for reproduction into an empty space (denoted by $\emptyset$) with a rate $q$, see e.g.~\cite{reichenbach_prl08,frey_pa10,szczesny_epl13}. The reaction given by Eq.~(\ref{dom1}) corresponds to one species dominating and displacing another (dominance-removal) with rate $p$, whereas in Eq.~(\ref{dom2}) there is a zero-sum dominance-replacement with rate $z$.
Recent experiments have shown that in addition to cyclic dominance,
some species can mutate. For instance,  {\it E.~coli} bacteria are known to mutate~\cite{kerr_n02}, and it was found that the side-blotched lizards {\it Uta stansburiana} are not only engaged in cyclic competition, but they can also undergo throat colour transformations~\cite{sinervo_hormonesbehaviour2000}.
It is therefore also instructive to assume that each species can mutate into one of the other species with a rate $\mu$~\cite{mobilia_jtb10,szczesny_epl13}, according to
\begin{eqnarray}
\label{mut}
A \stackrel{\mu}{\longrightarrow}
\begin{cases} B  \\
C \end{cases}
\text{,}
\quad
B \stackrel{\mu}{\longrightarrow}
\begin{cases} A  \\
C \end{cases}
\text{,}
\quad
C \stackrel{\mu}{\longrightarrow}
\begin{cases} A  \\
B \end{cases}
\text{.}
\end{eqnarray}
Additional evolutionary processes that should be specified  are those encoding the movement of individuals. However, these are irrelevant in the absence of spatial structure and will therefore be introduced in the next section.
In the mean field limit of population size $N\to \infty$, the dynamics of the system evolving according to Eqs.~(\ref{dom1})-(\ref{mut}) is aptly described by the following rate equations for the densities $a(t)$ of $A$'s, $b(t)$ of $B$'s and $c(t)$ of $C$'s:
\begin{eqnarray}
\label{RE}
\frac{d a}{d t} &=&a\left[q\rho_{\emptyset} + z(b-c) -pc\right]+\mu(b+c-2a)\nonumber\\
\frac{d b}{d t} &=&b\left[q\rho_{\emptyset} + z(c-a) -pa\right]+\mu(a+c-2b)\\
\frac{d c}{d t} &=&c\left[q\rho_{\emptyset} + z(a-b) -pb\right]+\mu(a+b-2c),\nonumber
\end{eqnarray}
where $\rho_{\emptyset}=1-(a+b+c)$  denotes the density of empty sites.
These rate equations admit a steady state ${\bf s}^*=(a^*,b^*,c^*)$  in which $A, B$ and $C$ coexist at the same density $a^*=b^*=c^*=\frac{q}{p+3q}$ encompass three types of oscillatory dynamics:
\begin{enumerate}
\item[(i)] When $p,q>0$, $z\geq 0$ and $\mu=0$, the Eqs.~(\ref{RE}) can be rewritten as the rate equations of the  May-Leonard model~\cite{may_siam75}. In this case, the above coexistence state is unstable and there are also three absorbing stationary states corresponding to the population being composed of  only $A$, $B$ or $C$. As in the May-Leonard model, the dynamics is characterized by {\it heteroclinic cycles} connecting these three absorbing states, with the population being composed almost exclusively of either $A$, $B$ or $C$, in turn. It is worth noting that the  heteroclinic cycles
are degenerate when $z=0$~\cite{may_siam75}.
\item[(ii)] When $p,q>0$, $z\geq 0$ and $\mu>0$,
a supercritical Hopf bifurcation arises at the value $\mu=\mu_H=pq/(6(p+3q))$ of the bifurcation parameter:
the coexistence steady state is a stable focus when $\mu>\mu_H$, while it is unstable
when $\mu<\mu_H$ and the dynamics is thus characterized by a stable  {\it limit cycle} of frequency
$\omega_H\approx \sqrt{3}q(p+2z)/[2(p+3q)]$~\cite{szczesny_epl13,mobilia_jtb10}.
\item[(iii)] When $z> 0$ and $p=q=\mu=0$, the state $\emptyset$ plays no role in the dynamics and one recovers the cyclic Lotka-volterra model~\cite{frachebourg_pre96,frachebourg_jpa98}. The sum and product of the densities, $a(t)+b(t)+c(t)$ and $a(t)b(t)c(t)$, are conserved by (\ref{RE}) which results in nested {\it neutrally stable closed orbits} around the  coexistence fixed point which is a neutrally stable center~\cite{hofbauer_88}. It is worth noting that the cyclic dominance-replacement scheme with
asymmetric rates $AB \stackrel{z_A}{\longrightarrow} AA\,,\quad BC \stackrel{z_B}{\longrightarrow} BB\,,\quad CA \stackrel{z_C}{\longrightarrow} CC$
is also characterized by
a marginally stable coexistence fixed point ${\bf s}^*=\frac{1}{z_A+z_B+z_C}(z_B,z_C,z_A)$ surrounded
by neutrally stable orbits on which the mean field dynamics takes place,
see Fig.~\ref{MF_portrait}~\cite{reichenbach_pre06,he_q_pre10}.
\end{enumerate}
The evolution in well-mixed populations of finite size,
$N <\infty$ is usually described in terms of  birth-and-death Markov chains, see e.g.~\cite{gardiner_04}.
In this case, the interactions between a finite number of
discrete individuals lead to fluctuations that may drastically affect the dynamics.
In particular, in the presence of absorbing states,
fluctuations are responsible for the extinction of two
strategies and the fixation of the surviving one, see Fig.~\ref{MF_portrait}~\cite{reichenbach_pre06,he_q_pre10}.
Questions concerning the mean time necessary for two species to go extinct and
the probability that one specific strategy survives
have recently attracted significant attention and were mostly addressed
by means of stochastic simulations. These are often efficiently carried out using the Gillespie
 algorithm~\cite{gillespie_1976}: for instance,
cyclic competition and reproduction according to Eqs.~(\ref{dom1})-(\ref{rep}) are implemented by
allowing at each time increment a reproduction event with probability
$q/(p+q+z)$, a dominance displacement move with probability
$p/(p+q+z)$ and a dominance replacement update with probability
$z/(p+q+z)$. When $p>0$ and $\mu=0$ and the mean field dynamics is characterized by
heteroclinic cycles, the mean extinction time has been found to be of order ${\cal O}(N)$. The existence of quantities conserved by the mean field dynamics
allows some analytical progress in the case of cyclic dominance with replacement ($p=q=\mu=0$)
for which it was shown that two species die out after a mean time that scales linearly
with the system size $N$~\cite{reichenbach_pre06, parker_pre09, dobrinevski_pre12}.
In this class of models with asymmetric rates, it was also shown the survival probability
follows an intriguing ``law of the weakest": the strategy with the smallest dominance rate is the most likely to survive and to fixate the population~\cite{frean_prsb01, berr_prl09}, as illustrated in Fig.~\ref{MF_portrait}. Some other aspects of extinction in RPS games in finite and well-mixed populations have been considered in Ref.~\cite{traulsen_prl05, claussen_prl08} and studied in further details in Ref.~\cite{galla_jtb11}.

\section{Rock-paper-scissors game in structured populations}
\label{rps}

In a structured population, or alternatively, in a spatial system, the interactions between players are defined by an interaction network and are thus limited. Because players have only limited interactions with their neighbors, a strategy (or species) will not necessarily meet its superior (or predator). This restriction has far-reaching consequences, because it enables pattern formation, and through that the survival of all three competing strategies with a time-dependent frequency akin to the Red Queen \cite{van-valen_et73, dieckman_jtb95, dercole_prsb10}. For example, it is possible to observe propagating fronts and spiral waves between the competing strategies \cite{szabo_pre99, peltomaki_pre08, szczesny_epl13}, which are particularly fascinating and complex if the invasion rates between different pairs of strategies are significantly different \cite{juul_pre13}.  In general, self-sustained oscillations of all three strategies are possible in spatially homogeneous populations \cite{neumann_g_dcdssb07, neumann_g_jmb07}
and result from periodic dynamics around a long-lived metastable  coexistence state
whose life time is determined by the system size (or the size of the population) \cite{muller_pre10,he_q_epjb11}. Hence, if the system size is sufficiently large all possible combinations of invasion fronts appear, resulting in a time-independent constant frequency of strategies, depending only on the invasion rates between the three competing strategies.

In comparison to a well-mixed population, a spatial system allows us to introduce an additional microscopic process that has no effect if everybody interacts with everybody else, namely site exchange between strategies or between an individual player and an empty site. This gives rise to mobility, which arguably has a profound impact on the outcome of the RPS game as it can both promote and impede biodiversity \cite{reichenbach_n07}. In contrast to well-mixed systems, where as seen intrinsic noise can jeopardize the coexistence of species \cite{reichenbach_pre06}, depending on the level of mobility, species coexistence can be more easily maintained  than under well-mixed conditions~ \cite{reichenbach_n07,reichenbach_prl07} and stochastic effects can have counterintuitive effects at some bifurcation point be reduced by stable patterns in spatial systems \cite{reichenbach_prl08}.

The mentioned mixing of strategies (directly or via empty sites) can also influence the cyclic competition in aquatic media \cite{karolyi_jtb05}. Furthermore, the degree of mixing determines the way of extinction, like the annihilation of domains, heteroclinic orbits, or traveling waves in a finite system \cite{rulands_jsm11}. The RPS game can also be regarded as a cyclical Lotka-Volterra model, and indeed it has been shown that a two-species Lotka-Volterra system with empty cites can give rise to global oscillations \cite{provata_pre03}. In general, however, it is important to acknowledge that the structure of the interaction network and mobility can both have a significant impact on the evolutionary outcome of cyclical interactions, and in the following subsections we review this in more detail, as well as survey the usefulness of the complex Ginzburg-Landau equation.

\subsection{Interaction networks}
\label{topology}

In a structured population species should contact in real-space and their interaction is characterized by an appropriate graph. Lattices are the simplest among these interaction networks, where every player has the same number of neighbors. By far the most widely used is the square lattice, although alternatives such as the honeycomb lattice, the triangular lattice, and the kagom{\'e} lattice are frequently employed as well to study the relevance of the degree of players and the role of clustering. While the square and the honeycomb lattice both have the clustering coefficient ${\cal C}$ equal to zero, the kagom{\'e} and the triangular lattice both feature percolating overlapping triangles, such that their clustering coefficient is ${\cal C}=1/3$ and ${\cal C}=2/5$, respectively. In general, lattices can be regarded as an even field for all competing strategies where the possibility of network reciprocity is given \cite{nowak_n92b}. Furthermore, as there are many different types of lattices, it is possible to focus on very specific properties of cyclical interactions and test what is their role in the evolutionary process. In some cases, like by the competition of bacteria \cite{kerr_n02, kim_pnas08, prado_evol08, frey_pa10, nahum_pnas11} or for the description of parasite-grass-forb interactions \cite{cameron_jecol09}, lattices are also an apt approximation of the actual interaction network. In more complex systems, however, the description of interactions requires the usage more intricate networks, which typically have broad-scale degree distribution and small-world properties \cite{albert_rmp02, boccaletti_pr06}. The question how the topology of the interaction network affects the outcome of cyclical interaction, and game theoretical models in general \cite{szabo_pr07, roca_plr09, perc_bs10, perc_jrsi13}, is one of great interest.

\begin{figure}
\centerline{\epsfig{file=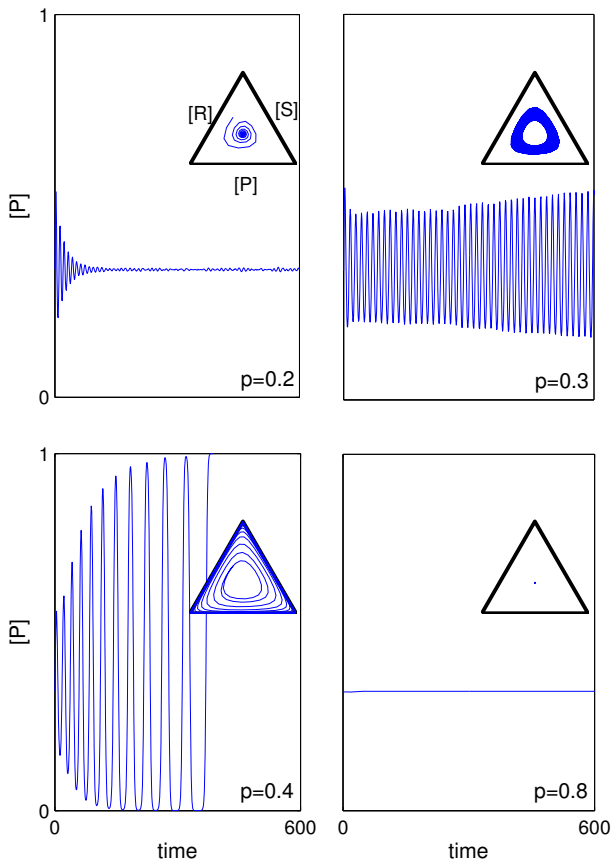,width=8cm}}
\caption{Transitions between ``stationary $\to$ oscillatory $\to$ absorbing $\to$ stationary'' phases in a coevolutionary RPS game where an invaded player can severe the link towards its predator with a probability $P$. For further details we refer to \cite{demirel_epjb11}. The reported nonequilibrium phase transitions are qualitatively similar as those observed on interaction networks with annealed randomness \cite{szolnoki_pre04b}, as well as those observed in coevolutionary social dilemma games \cite{szolnoki_epl09, szolnoki_njp09}. The emergence of oscillatory behavior akin to the Red Queen thus appears to be a robust phenomenon on coevolving networks.}
\label{demirel}
\end{figure}

As an initial departure from lattices towards more complex interaction topologies, it is possible to consider networks with dispersed degrees. As demonstrated by Masuda and Konno, the resulting heterogeneity can help to maintain the stable coexistence of species in cyclic competition \cite{masuda_pre06}. Another possibility is to introduce permanent shortcut links that connect distant players, thereby effectively introducing small-world properties. In contrast to lattices, on small-world networks the frequencies of competing strategies oscillate even in the large system size limit, as shown \cite{szolnoki_pre04b,szabo_jpa04,sun_rs_cpl09,rulquin_pre14}. The explanation behind this observation lies in the fact that the introduction of shortcuts introduces quenched randomness into the interaction network. If the magnitude of this randomness exceeds a threshold value, i.e., if the number of shortcut links is sufficiently large, the amplitude of frequency oscillations can be so large that the system will always terminate into an absorbing state during the evolutionary process. These results lead to the conclusion that not only the limited interaction range that is imposed by structured populations, but even more so the explicit spatial distribution of competing species is the crucial property that determines coexistence and biodiversity \cite{laird_oikos14}. This observation becomes particularly interesting if we compare it with the observations made in the realm of spatial social dilemma games, where the limited interaction range was found to be the decisive factor for the promotion of cooperation based on network reciprocity \cite{nowak_s06, szabo_pr07}.

Properties that are typical of small worlds can be introduced not only by means of permanent shortcut links between distant players, but also by means of temporary long-range links that replace nearest-neighbor links with probability $P$. This process is typically referred to as annealed randomness \cite{szolnoki_pre04b}. But while topological properties of annealed and quenched small worlds may be very similar, there is a relevant difference in the evolutionary outcomes of cyclical interactions taking place on them. Namely, annealed randomness no longer gives rise to oscillatory frequencies of the three competing strategies in the RPS
game. For sufficiently large values of $P$, the oscillations are replaced by steady state behavior \cite{ying_cy_jpa07}, as is commonly observed on regular lattices. Interestingly, similar results were also reported for a four-strategy cyclic dominance game in \cite{zhang_gy_pre09}. But as we will review later, the transitions between stationary and oscillatory states in cyclical interactions can in fact be driven by many different mechanisms, even if the interaction network thereby remains unchanged.

The topology of the interaction network may also change as a result of a particular coevolutionary rule, as studied frequently before in the realm of social dilemma games. Note that when not only individual strategy but also other player-specific character (such as connectivity, strategy adopting- or passing capacity, etc.) may change in time then we can observe parallel evolutions, co-evolution of strategies and other quantities, see \cite{perc_bs10} for a review. In particular, rewiring could be the consequence of invasions between strategies \cite{szolnoki_epl09, szolnoki_njp09}. Although games of cyclic dominance received comparatively little attention in this respect, there is a study by Demirel et al.~\cite{demirel_epjb11}, which builds exactly on this premise. Namely, the loser of each particular instance of the game can either adopt the winning strategy (an invasion occurs), or it can rewire the link that brought the defeat. It was shown that nonequilibrium phase transitions occur as a function of the
rewiring strength, as illustrated in Fig.~\ref{demirel}. The observation of the ``stationary $\to$ oscillatory $\to$ absorbing $\to$ stationary'' phase transitions is conceptually similar to those observed in the realm of interaction networks with annealed randomness. In general, it is thus possible to conclude that oscillatory behavior in coevolving networks is a robust phenomenon --- a conclusion which is further corroborated by agent-based simulations showing that coevolving networks can influence the cycling of host sociality and pathogen virulence \cite{prado_jtb09}.

\subsection{Mobility}
\label{mobile}

As briefly explained above, besides the properties of the interaction network, mobility, as is the case in general for evolutionary game theoretical models \cite{vainstein_jtb07, meloni_pre09, sicardi_jtb09, cardillo_pre12, vainstein_pa14}, can also play a decisive role in the maintenance of biodiversity in games of cyclic dominance. The consideration of mobility has both theoretical and practical aspects. Theoretically, the question is whether mobility affects the stability of spatial patterns that typically emerge in structured populations, and if yes, how? From the practical point of view, it is important to acknowledge the fact that in real life prey and predators are frequently on the move in order to maximize their chances of survival. Accordingly, their neighborhoods change, which is in agreement with the effect that is brought about by mobility in simulations. The crucial point to clarify is whether mobility simply shifts the evolutionary outcomes towards those observed in well-mixed populations, or whether there emerge unexpected phenomena that disagree with mean-field predictions. As we will review in what follows, the introduction of mobility in structured populations is certainly much more than simply a transition towards well-mixed conditions.

\begin{figure}
\centerline{\epsfig{file=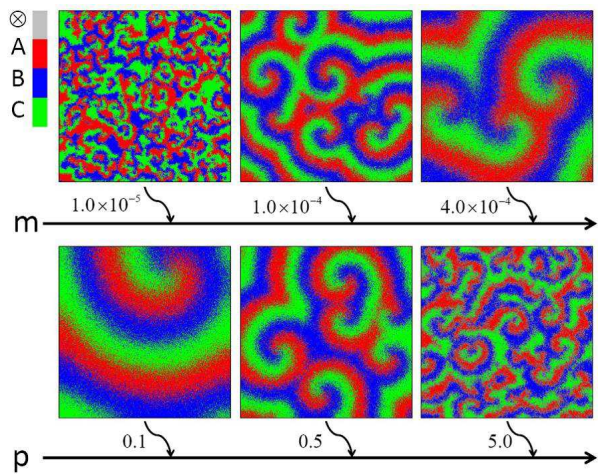,width=8cm}}
\caption{Typical spatial patterns, which emerge spontaneously from random initial conditions on a square lattice. Top panel illustrates how the characteristic length scale of spiral waves increases with increasing mobility $m$ from left to right. Here the competition and reproduction rate are both $p=q=1$. Bottom panel shows the reverse transition, where the characteristic length scale of spiral waves decreases as the competition rate $p$ increases from left to right. Here the mobility $m=10^{-4}$ and the reproduction rate $q=1.0$ are fixed. Colors red, blue and green denote rock (species $A$), paper (species $B$) and scissor (species $C$), respectively. Empty sites are gray and denoted by $\bigotimes$. The system size is $N=512^2$. For further simulation details we refer to \cite{jiang_ll_pre11}.}
\label{mobandcom}
\end{figure}

Pioneering work on the subject has been done by Reichenbach et al.~\cite{reichenbach_n07, reichenbach_jtb08}, who showed that mobility (or migration) has a critical impact on biodiversity. In particular, small mobility that is below a critical threshold promotes species coexistence, while large mobility that is above the threshold destroys biodiversity. Behind these observations are spontaneously emerging spiral patterns, which in general sustain the coexistence of all three competing strategies \cite{szabo_pre99, szolnoki_pre04}. Under the influence of mobility, however, the spirals grow in size, and above the critical threshold, they outgrow the system size causing the loss of biodiversity. Accordingly, in \cite{reichenbach_n07} it was thus argued that mobility can both promote and jeopardize biodiversity in RPS
games.

As most studies on this subject, in this section we will focus on RPS games on two-dimensional square lattices of total size $N$ (here, linear size is $\sqrt{N}$). The population's composition changes according to the dominance-removal reactions given by Eq.~(\ref{dom1}) and reproduction reactions given by Eq.~(\ref{rep})
(i.e. $q,p>0$ and $z=\mu=0$, which yields degenerate heteroclinic cycles at mean field level; see Sec.~\ref{non-spatial}). Furthermore, it is assumed that individuals can move by swapping their position with any neighbor at rate
$\gamma$ according to the scheme
\begin{eqnarray}
\label{exch}
XY &\stackrel{\gamma}{\longrightarrow}& YX \nonumber\\
X\emptyset &\stackrel{\gamma}{\longrightarrow}& \emptyset X,
\end{eqnarray}
where $X,Y\in (A,B,C)$.
In many references, mobility is measured in continuum limit by introducing the mobility or
diffusion coefficient $m=2\gamma/N$ where $N$ is the system size, see e.g.~\cite{reichenbach_n07}.

To focus on the effects of mobility $m$ on pattern formation and biodiversity, the rate of competition $p$ and reproduction $q$ can initially be set fixed at $1$. Starting from random initial conditions where each lattice site is with equal probability either occupied by rock $A$, paper $B$, scissor $C$ or left empty, we find that the emerging spatial patterns depend sensitively on mobility. As shown in the top row of Fig.~\ref{mobandcom}, as mobility increases from left to right, the characteristic scale of spiral waves increases as well. It was found that at a critical value of mobility, $m=m_{c}=(4.5\pm 0.5)\times 10^{-4}$ for $p=q=1$, the length scale of the spiralling patterns outgrows the system size resulting in two species dying out and the dynamics settling in the absorbing of the remaining species~\cite{reichenbach_prl07, reichenbach_prl08, peltomaki_pre08, lamouroux_pre12}. The diversity is lost, whereby the extinction probability is defined as the disappearance of two species after a waiting time that is proportional with the system size. As the system size increases, this probability approaches zero if $m < m_{c}$, and it tends to $1$ if $m > m_{c}$.

\begin{figure}
\centerline{\epsfig{file=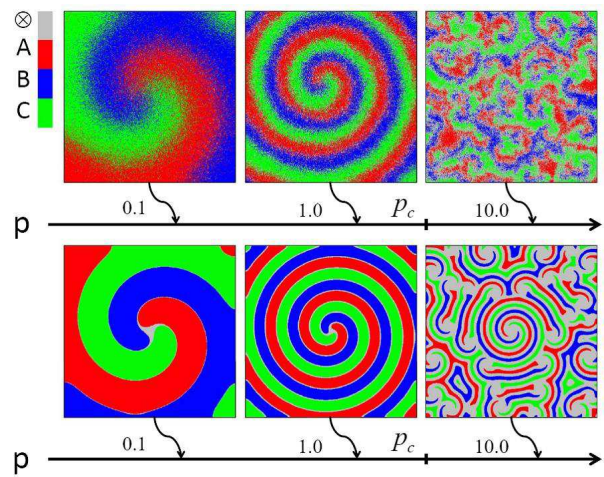,width=8cm}}
\caption{Disintegration of single-armed spiral waves as the competition rate $p$ increases past the critical values $p_c$. Depicted are typical spatial patterns that emerge from a prepared initial state (see Fig.~2 in \cite{jiang_ll_pre11} for details) for different values of $p$ (see legend). The top row depicts results obtained with Monte Carlo simulations for $m = 5.0 \times 10^{-5}$ and $q=1$, while the bottom row depicts results obtained with partial differential equations describing the spatial RPS game (see Eq.~(\ref{PDE})) for $D_m = 5.0 \times 10^{-5}$ and $q=1$. Disordered spatial patterns emerge as soon as $p>p_c=2.3$. Colors red, blue and green denote rock (species $A$), paper (species $B$) and scissor (species $C$), respectively. Empty sites are gray and denoted by $\bigotimes$. For further simulation details we refer to \cite{jiang_ll_pre11}.}
\label{spirals}
\end{figure}

In contrast to mobility, competition promotes coexistence of species because the typical length of the spatial patterns decreases with increasing competition rate $p$. This is illustrated in the bottom row of Fig.~\ref{mobandcom} from left to right. Macroscopic spiral waves emerge at $p=0.1$, but they become smaller and fragmented as $p$ increases. The fragmentation is due to the fact that large competition rates introduce empty sites around small patches that contain all three competing strategies (i.e. at the spiral core), which in turn leads to the disintegration of macroscopic spirals typically observed away from the core. The diversity is therefore favored at larger values of $p$, as they effectively prevent the spirals outgrowing the system size. Similar phenomenon can also be observed in the generic metapopulation model. In the latter case, however, the fragmentation is a consequence of the meeting of spiral fronts, as illustrated Fig.~\ref{ssa_different_mobilities}~\cite{szczesny_epl13,szczesny_PhD}.

Since pattern formation obviously plays a critical role in warranting the coexistence of the three competing strategies, it is particularly important to understand how the spatial patterns evolve. Prepared initial conditions are strongly recommended in such cases, as they can help  identify critical processes that may either enhance or destroy the stability of coexistence. The emergence of spirals can be engineered if the evolution is initiated from three roundish areas, each holding a particular strategy, while the rest of the lattice is empty \cite{jiang_ll_lnics09}. Based on such a procedure, it has been shown that there exists a critical competition rate $p_{c}$ below which single-armed spirals in finite populations are stable. For $p>p_{c}$ the spirals break up and form disordered spatial structures because too many empty sites allow the emergence of competing mini spirals \cite{jiang_ll_pre11}. This transition is demonstrated in the top row of Fig.~\ref{spirals} from left to right.

\begin{figure}
\centerline{\epsfig{file=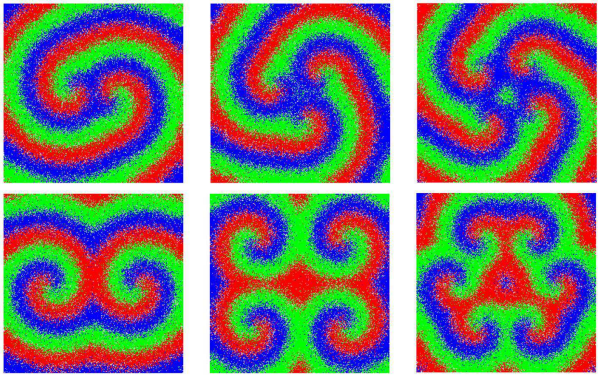,width=8cm}}
\caption{Beautiful examples  of multi-armed spirals with two, three and four arms (top row from left to right), as well as anti-spirals with one, two, and three pairs (bottom row from left to right panels). Special initial conditions have been used, which may provide further insight with regards to the robustness of biodiversity against mobility. Colors red, blue, green and gray denote rock (species $A$), paper (species $B$), scissor (species $C$) and empty sites, respectively, like in Figs.~\ref{mobandcom} and \ref{spirals}. Here the mobility rate is $m=6.0\times10^{-5}$ and the system size is $N=512^2$. For further simulation details and the special conditions we refer to \cite{jiang_ll_pla12}.}
\label{multispirals}
\end{figure}

Although Monte Carlo simulations are applied most frequently for studying spatial evolutionary games (see reviews \cite{szabo_pr07} and \cite{perc_jrsi13} for details, and  \cite{szczesny_epl13} for the use of the spatial Gillespie algorithm~\cite{gillespie_1976}), when density fluctuations are neglected, the dynamics of the spatial RPS game on a domain of unit size can be described by the following of partial differential equations obtained by letting the lattice spacing $1/N$ vanish in the continuum limit $N\to \infty$:
\begin{eqnarray}
\begin{array}{lll}
\partial_t a(\mathbf{r},t)=D_m\nabla^2 a(\mathbf{r},t)&+&qa(\mathbf{r},t)\rho_{\emptyset}-pc(\mathbf{r},t)a(\mathbf{r},t)\\
\partial_t b(\mathbf{r},t)=D_m\nabla^2 b(\mathbf{r},t)&+&qb(\mathbf{r},t)\rho_{\emptyset}-pa(\mathbf{r},t)b(\mathbf{r},t)\\
\partial_t c(\mathbf{r},t)=D_m\nabla^2 c(\mathbf{r},t)&+&qc(\mathbf{r},t)\rho_{\emptyset}-pb(\mathbf{r},t)c(\mathbf{r},t),
\end{array}
\label{PDE}
\end{eqnarray}
where $\mathbf{r}$ labels the two-dimensional position, while
$\rho_{\emptyset}$ and $D_m$ denote the density of empty site and the diffusion coefficient (on a domain of unit size, $D_m$ coincides with $m$), respectively. This set of partial differential equations can be solved numerically by standard finite difference methods, and as illustrated in the bottom row of Fig.~\ref{spirals}, the results are similar to those obtained via Monte Carlo simulations.

\begin{figure}
\centerline{\epsfig{file=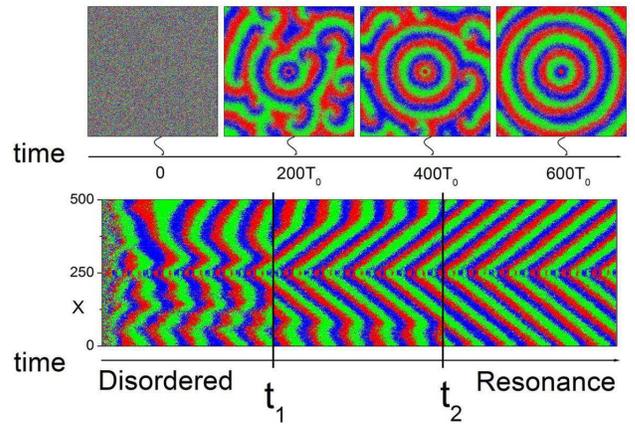,width=8.4cm}}
\caption{Emergence of target waves due to the application of a periodic current of all three strategies at the center of a square lattice. The top row shows characteristic snapshots of the spatial grid over time from left to right (note how the initial random state is gradually transformed into a global target wave). The bottom row shows the corresponding space-time plot, where the emergence of the cone in the traveling front indicates the emergence of resonance between the periodic current and the global oscillations. Colors red, blue, green and gray denote rock (species $A$), paper (species $B$), scissor (species $C$) and empty sites, respectively, like in Figs.~\ref{mobandcom} and \ref{spirals}. The system size is $N=512^2$. For further details we refer to \cite{jiang_ll_njp09}.}
\label{target}
\end{figure}

By exploiting further the potential of prepared initial conditions, it is also possible to observe multi-armed spirals, as well as pairs of anti-spirals with a finite number of arms and pairs \cite{jiang_ll_pla12}. We show examples of this fascinating pattern formation in Fig.~\ref{multispirals}. An in-depth qualitative analysis reveals that such multi-armed spirals and anti-spirals are more robust to mobility that single-armed spirals, persisting not only at small but also at intermediate mobility rates. In addition to prepared initial conditions, periodic currents (or localized forcing) can also be applied to evoke specific spatial patterns. If a periodic current of all three strategies is applied at the center of the lattice, target waves can be observed, as reported in \cite{jiang_ll_njp09}. Such phenomena emerge due to the interplay between local and global system dynamics, in particular the intermittent synchronization between the periodic current and the global oscillations of the density of the three competing strategies on the spatial grid. In Fig.~\ref{target} we show how the global target wave (top row from left to right) emerges due to the emergence of resonance between the periodic current and the global oscillations (bottom row from left to right).

The impact of mobility can also be studied in off-lattice simulations, where the radius of the interaction range crucially affects coexistence.
In particular, a transition from coexistence to extinction is uncovered with a strikingly non-monotonous behavior in the coexistence probability \cite{ni_x_pre10}.
Close to the minimal value of the coexistence probability and under the influence of intermediate mobility, it is even possible to observe transitions between spiral and plane waves over time, as depicted in Fig.~2 of \cite{ni_x_pre10}. In general, strong mobility can either promote or impair coexistence depending on the interaction range, although diversity is quite robust across large regions of the parameter space \cite{ni_x_c10}. Importantly, these phenomena are absent in any lattice-based model, and it is possible to explain them using partial differential equations similar to Eq.~(\ref{PDE}). Off-lattice simulations of RPS and related evolutionary games are still relatively unexplored. Partly this is surely due to significantly higher computational resources that are needed to simulate such systems in comparison to traditional spatial games, yet on the other hand, the efforts may be well worth while as the off-lattice setup is closer to some actual conditions than lattice-based simulations.

Another interesting phenomenon to consider that is directly related to mobility is epidemic spreading. The effects of epidemic spreading on species coexistence in a spatial RPS
game have been studied in \cite{shi_hj_pre10, wang_wx_pre10b, wang_wx_pre11}. It has been shown that in the absence of epidemic spreading, there exist extinction basins that are spirally entangled for $m>m_{c}$, while basins of coexistence emerge for $m<m_{c}$. Moreover, it has been discovered that intra-species infection promotes coexistence, while inter-species spreading fails to evoke the same effect \cite{shi_hj_pre10, wang_wx_pre10b, wang_wx_pre11}.

To close this subsection, we note that many more interesting and counterintuitive phenomena are observable when more than three strategies compete for space. In a four-strategy system, for example, when a two-strategy neutral alliance may emerge, the intensity of mobility is able to directly select the winning solution of the subsystem \cite{dobrinevski_pre14}. More specifically, if the mobility is small the RPS-type cyclic dynamics is dominant, while if the mobility is large the two-strategy neutral alliances take over. Further fascinating examples related to cyclical interactions between more than three strategies are reviewed in Section~\ref{four}.

\subsection{Metapopulation and nonlinear mobility}
\label{meta-nonlinmob}
 In the previous section, as in the vast part of the literature,
the two-dimensional RPS models were implemented  on square
lattices whose nodes were either empty or occupied by an $A,B$ or $C$ player,
and the interactions described by Eqs.~(\ref{dom1}),(\ref{rep}),(\ref{exch}) involved agents on
neighboring sites.
Inspired by the experiments of \cite{kerr_n02,kerr_n06,nahum_pnas11},
an alternative metapopulation modeling approach allowing further analytical progress has recently been put forward, see e.g. \cite{szczesny_epl13,rulands_pre13,lamouroux_pre12}.
In the metapopulation formulation, the lattice consists
of an array of $L\times L$ patches each of which comprises a well-mixed (sub-)population
of constant size $N$ (playing the role of the carrying capacity)
that consists of $A,B$ or $C$ individuals and empty spaces.
The RPS game is then implemented according to the general reactions
Eqs.~(\ref{dom1})-(\ref{mut})
between individuals within the same patch (intra-patch reactions),
 while the spatial degrees of freedom are granted
 by allowing the individuals to move
between neighboring patches. At this point, it is interesting to
divorce the pair-exchange (with rate $\gamma_e$)
from the hopping (with rate $\gamma_d$) process,
according to the scheme ($X,Y= A,B,C$)
\begin{eqnarray}
\label{nonlinmob}
XY &\stackrel{\gamma_e}{\longrightarrow}& YX \nonumber\\
X\emptyset &\stackrel{\gamma_d}{\longrightarrow}& \emptyset X.
\end{eqnarray}
As shown in \cite{szczesny_epl13,SMR_axv14}, this leads to {\it nonlinear mobility} when $\gamma_e\neq \gamma_d$
(see also \cite{he_q_pre10,he_q_epjb11}).
In biology, organisms are in fact known not to simply move diffusively, but to sense and respond to their environment. Here, Eq.~(\ref{nonlinmob}) allows us to discriminate between the movement in crowded regions, where mobility is dominated by pair-exchange, and mobility in diluted regions where hopping can be more efficient.

The generic two-dimensional metapopulation model
of Eqs.~(\ref{dom1})-(\ref{mut}), supplemented by
 Eq.~(\ref{nonlinmob}),
is ideally-suited to capture stochastic effects via size expansion in $N$~\cite{gardiner_04}
and, to lowest order (where all fluctuations are neglected) and in the continuum limit~\cite{SMR_axv14}, yields the following partial differential equations (PDEs)
\begin{eqnarray}
\label{PDEmeta}
\partial_t a&=&
\gamma_d\nabla^2 a+\gamma_{{\rm nl}}[a\nabla^2(b+c)-(b+c)\nabla^2 a]\nonumber\\
&+&a\left[q\rho_{\emptyset} + z(b-c) -pc\right]+\mu(b+c-2a) \nonumber\\
\partial_t b&=&\gamma_d\nabla^2 b+\gamma_{{\rm nl}}[b\nabla^2(a+c)-(a+c)\nabla^2 b]\nonumber\\
&+&b\left[q\rho_{\emptyset} + z(c-a) -pa\right]+\mu(a+c-2b)
\\
\partial_t c&=&\gamma_d\nabla^2 c+\gamma_{{\rm nl}}[c\nabla^2(a+b)-(a+b)\nabla^2 c]\nonumber\\
&+&c\left[q\rho_{\emptyset} + z(a-b) -pb\right]+\mu(a+b-2c), \nonumber
\end{eqnarray}
with  $\gamma_{{\rm nl}}=\gamma_d-\gamma_e$ and the notation
$a\equiv a({\bf r},t)$,  $b\equiv b({\bf r},t)$ and  $c\equiv c({\bf r},t)$. We notice
that nonlinear diffusive terms of the form  $a\nabla^2(b+c)-(b+c)\nabla^2 a$
arise when $\gamma_{{\rm nl}}\neq 0$.
In \cite{szczesny_epl13,szczesny_PhD,SMR_axv14}, the solutions of
Eq.~(\ref{PDEmeta})
have been shown shown to accurately match the stochastic simulations of the metapopulation model
performed using a spatial version of the Gillespie algorithm~\cite{gillespie_1976}.
	\begin{figure}
\centerline{\epsfig{file=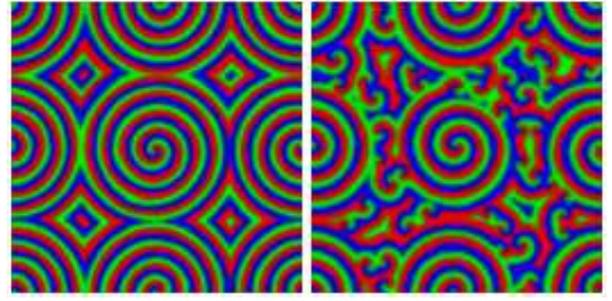,width=8cm}}
		\caption{Influence of nonlinear mobility on spiralling patterns  in the metapopulation RPS model defined by (\ref{dom1})-(\ref{mut}) and (\ref{nonlinmob}) with $(\gamma_d,\gamma_e)=(0.05,0.05), (0.20,0.05)$ from left to right respectively:
Typical snapshots in lattice simulations starting from ordered initial conditions
and with parameters $L^2=128^2$, $N=1024$, $p=q=1$, $z=0.1$ and $\mu=10^{-6} \ll \mu_H=0.042$ (far below the Hopf bifurcation point). Left panel: Nonlinear mobility  enhances the convective instability and  far field breakup of spiral waves. See \cite{szczesny_epl13} and movies~\cite{szczesny_movies} for further details.}
		\label{ssa_different_mobilities}
	\end{figure}
It has been shown that the PDEs (\ref{PDEmeta}) accurately capture the properties of the
lattice metapopulation model as soon as the carrying capacity is $N\gtrsim 64$,  while
their derivation assumes $N\gg 1$~\cite{szczesny_epl13,SMR_axv14}. Quite remarkably, it turns out that certain
outcomes of  stochastic simulations with $N=2-16$ are qualitatively reproduced by the solutions of Eqs.~(\ref{PDEmeta})~\cite{szczesny_movies}.

In the next section we shall discuss how the PDEs (\ref{PDEmeta})  can be used to derive a  Complex Ginzburg-Landau equation (CGLE), with an effective linear diffusive term, from which many of the spatio-temporal properties of the spiraling patterns that we have encountered can be predicted. Yet, some intriguing features of the model triggered by nonlinear diffusion  cannot be captured by the CGLE. An interesting effect of nonlinear mobility
appears in the RPS model of Eqs.~(\ref{dom1})-(\ref{mut}) at low mutation rates when $\gamma_d>\gamma_e$: as illustrated in Fig.~\ref{ssa_different_mobilities}, in this case the nonlinear mobility promotes the far field breakup of spiral waves and enhances their convective instability~\cite{szczesny_epl13,szczesny_movies,SMR_axv14}.

\subsection{Complex Ginzburg-Landau equation}
\label{cgle}
The CGLE is a celebrated and well-studied nonlinear equation
commonly used in the physics community to describe a variety of phenomena spanning from superconductivity, superfluidity, liquid crystals to field theory~\cite{aranson_rmp02}. After the introduction of a phenomenological time-dependent Ginzburg-Landau theory for superconductors~\cite{schmid_66}, the CGLE was derived
in the context of fluid dynamics~\cite{stewartson_jmf71,diprima_jmf71} and is known for its ability to predict the emergence of complex coherent structures, such as spiral waves in two-spatial dimensions, as well as chaotic behavior. Its
 fascinating properties  have received considerable interest and are the subject of numerous reviews, see e.g.~\cite{vanSaarloos_physica_D_92,aranson_rmp02}. It has recently been realized that
many of the spatio-temporal properties of the two-dimensional RPS models can be understood by means of a suitable mapping onto a CGLE~\cite{reichenbach_n07,reichenbach_prl07,reichenbach_jtb08}.

The first attempts to use the CGLE in the context of spatial RPS games have been carried out in
\cite{reichenbach_n07,reichenbach_prl07,reichenbach_jtb08} for the model with dominance-removal
reactions (\ref{dom1}), while $z=\mu=0$, and pair exchange in Eq.~(\ref{exch}). At mean field level, this model is characterized by an unstable coexistence fixed point ${\bf s}^*$ with one stable eigenvector,
and the dynamics quickly approaches a manifold
tangent to the plane normal to the stable eigenvector of ${\bf s}^*$. In
\cite{reichenbach_n07,reichenbach_prl07,reichenbach_jtb08}
that manifold was computed  to quadratic
order around ${\bf s}^*$.
On such manifold,
the dynamics becomes two-dimensional and the
mean field flows approach the absorbing boundaries
of the phase portrait where they linger and form a heteroclinic cycle~\cite{may_siam75,hofbauer_98}.
 Exploiting a suitable nonlinear (``near-identity'') transformation, effectively mapping $(a,b,c)-{\bf s}^*$ onto the new variables ${\bf z}=(z_1,z_2)$ and by retaining terms up to cubic order in ${\bf z}$, the rate equations (\ref{RE}) was then recast in the normal form of a
 Hopf bifurcation. As explained in \cite{reichenbach_jtb08,frey_pa10}, where details of the derivation can be found, space and movement were finally reinstated by adding a linear diffusive term to rate equations
in the ${\bf z}$ variables yielding a CGLE.
It has to be noted that such an approach relies on  three uncontrolled approximations:
\\
(i) It consists of approximating heteroclinic cycles by stable limit cycles resulting from a fictitious Hopf bifurcation.
\\
(ii) While the new variables $z_1$ and $z_2$ are assumed to be small quantities, and therefore the mapping is expected to be suitable near ${\bf s}^*$, its use is not restricted to the vicinity of ${\bf s}^*$.
\\ (iii) Even though mobility is mediated by  pair exchange
in Eq.~(\ref{exch}), which results in  linear diffusion in the PDEs~(\ref{PDE}), the nonlinear mapping
$(a,b,c)-{\bf s}^* \to {\bf z}$ generates {\it nonlinear diffusive terms} that are
ignored by the CGLE.

Notwithstanding, this approach has been able to explain many qualitative
features of the above model and, upon adjusting one fitting parameter, even obtain
approximations for the wavelength of spiraling patterns and the velocity of the propagating fronts~\cite{reichenbach_n07,reichenbach_prl07,reichenbach_jtb08}. This treatment was then extended to the case of dominance-removal and dominance-replacement cyclic competitions with linear mobility and no mutations ($p,z,\gamma>0, \mu=0$)~\cite{reichenbach_prl08,rulands_pre13}. Recently, a class of RPS-like models with more than three species (with dominance-removal and hopping but no mutations) has also been considered
by generalizing this approach~\cite{mowlaei_jpa14}.

Recently, an alternative analytical treatment has been devised to derive {\it perturbatively} the CGLE associated with the generic class of RPS models
defined by the reactions Eqs.~(\ref{dom1})-(\ref{mut}), and with mobility mediated by pair-exchange and hopping
in Eq.~(\ref{nonlinmob})~\cite{szczesny_epl13,SMR_axv14}.
This alternative derivation has been carried out within the metapopulation formulation outlined in Sec.~\ref{meta-nonlinmob}. The key observation is that the mean field dynamics of the model defined by Eqs.~(\ref{dom1})-(\ref{mut}) is characterized by a
Hopf bifurcation at $\mu=\mu_H$ (see Sec.~\ref{non-spatial}) and by stable limit cycles when $\mu<\mu_H$.
A multiscale expansion around the Hopf bifurcation is appropriate
to derive a CGLE that aptly describes the fascinating
oscillatory  patterns arising on two-dimensional lattices when $\mu<\mu_H$. As explained in \cite{szczesny_epl13,szczesny_PhD}, a space and time perturbation expansion in the parameter $\epsilon \propto \sqrt{\mu_H - \mu}$ is performed by introducing the ``slow variables''
$(T, {\bf R})=(\epsilon^2 t, \epsilon{\bf r})$ and by expanding the densities in powers of $\epsilon$, see e.g.~\cite{miller_06}.
Hence, after a suitable linear transformation $(a,b,c)-(a^*,b^*,c^*) \to (u,v,w)$,
where $w$ decouples from $u$ and $v$ (to linear order),
one writes
$u({\bf r},t) = \sum_{n=1}^{3} \epsilon^n U^{(n)}(t,T,{\bf R})$,
 $v=\sum_{n=1}^{3} \epsilon^n V^{(n)}$ and $w= \sum_{n=1}^{3} \epsilon^n W^{(n)}$, where the functions $U^{(n)}, V^{(n)}, W^{(n)}$ are of order ${\cal O}(1)$.
Substituting these expressions into
Eq.~(\ref{PDEmeta})
with $U^{(1)}+iV^{(1)}=\mathcal{A}(T,{\bf R})e^{i\omega_H t}$, where $\mathcal{A}$ is a (complex) modulated amplitude,
the CGLE is thus derived by
imposing the removal of the secular term
arising at order $\mathcal{O}(\epsilon^3)$, which yields~\cite{szczesny_epl13,szczesny_PhD,SMR_axv14}
	\begin{equation}
	\label{CGLEepl}
		\partial_{T} \mathcal{A} =
		\delta \Delta_{\bf R} \mathcal{A} + \mathcal{A} - (1 + i c) |\mathcal{A}|^2 \mathcal{A},
	\end{equation}
where $\delta = \frac{3q\gamma_e + p\gamma_d}{3q+p},~\Delta_{\bf R}=\partial_{r_1}^2+\partial_{r_2}^2$ and, after rescaling $\mathcal{A}$ by a constant~\cite{szczesny_PhD},
	\begin{equation}
	\label{c3}
		c = \frac
		{12 z (6 q - p)(p + q) + p^2 (24q - p)}
		{3\sqrt{3} p (6q + p)(p + 2z)}.
	\end{equation}
It is worth stressing that the CGLE (\ref{CGLEepl}) is a controlled approximation
of the PDEs (\ref{PDEmeta}) around
the Hopf bifurcation point, and it is noteworthy that it involves a linear (real) effective diffusion coefficient.
	\begin{figure}
\centerline{\epsfig{file=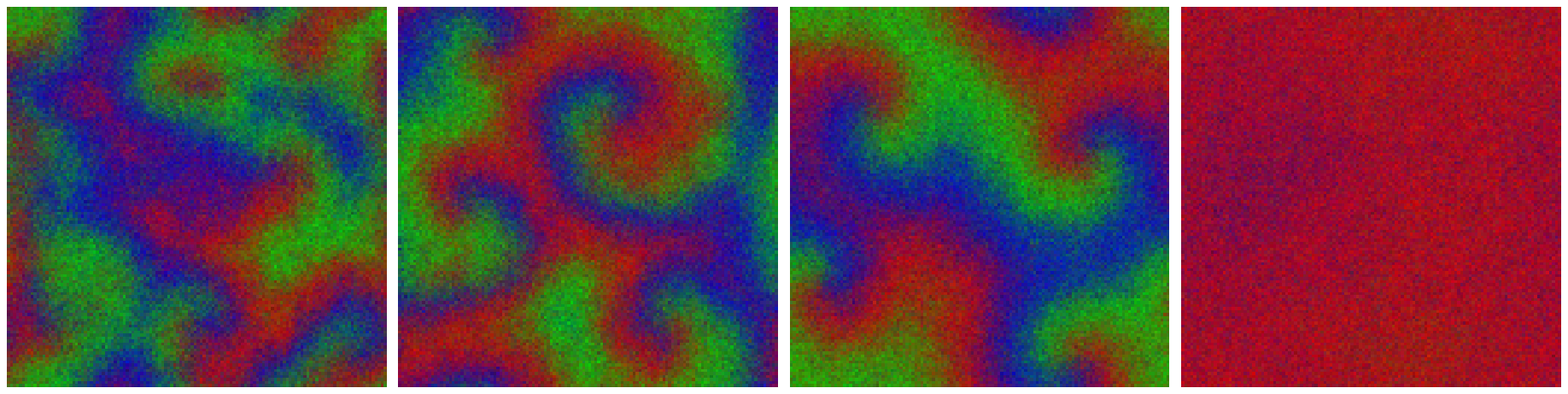,width=8.5cm}}\vspace{5 mm}
\centerline{\epsfig{file=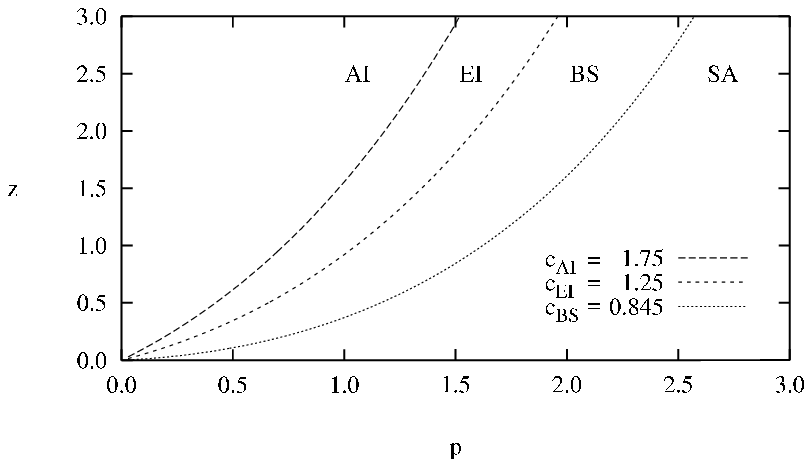,width=8.5cm}}
		\caption{Upper: Typical snapshots of the phases AI, EI, BS, SA (from left to right)
obtained from lattice simulations with parameters $q = \gamma_e=\gamma_d=1$,  $\mu=0.02$, $L=128, N=64$
and, from left to right, $z=(1.8, 1.2, 0.6, 0)$.
Lower: Phase diagram of the two-dimensional RPS system (\ref{dom1})-(\ref{mut}) around the
Hopf bifurcation point with contours of $c = (c_{{\rm AI}}, c_{{\rm EI}}, c_{{\rm BS}})$ in the $p-z$ plane, see text. We distinguish four phases: spiral waves are unstable in AI, EI and SA phases, while they are stable in BS phase. The boundaries between the phases have been obtained using (\ref{c3}). See \cite{szczesny_epl13} for full details.}
		\label{CGLE_phase_diagram}
	\end{figure}
The CGLE (\ref{CGLEepl}) permits an accurate characterization of the spatio-temporal patterns in the vicinity of the Hopf Bifurcation point. For simplicity, we here restrict $\sigma$ and $\zeta$ into [0,3]. Using the well-known phase diagram of the two-dimensional CGLE~\cite{aranson_pre93,aranson_rmp02}, one thus distinguishes four phases separated by the three critical values $(c_{{\rm AI}}, c_{{\rm EI}}, c_{{\rm BS}})\approx (1.75, 1.25, 0.845)$, as illustrated in Fig.~\ref{CGLE_phase_diagram} (upper panel):
the absolute instability (AI) phase in which no stable spiral waves cannot be sustained (when $c>c_{{\rm AI}}$), the Eckhaus instability (EI) phase in which spiral waves are convectively unstable, with spirals' arms that far from the cores first distort and then break up (when $c_{{\rm EI}}<c<c_{{\rm AI}}$), the bound state (BS) phase characterized by stable spiral waves with well defined wavelength and frequency (when $c_{{\rm BS}}<c<c_{{\rm EI}}$), and a phase in which spiral waves annihilate
when they collide (SA phase, when $0<c<c_{{\rm BS}}$).

The properties of the CGLE (\ref{CGLEepl}) have also be exploited to gain insight into the properties of the spatio-temporal patterns arising away from the Hopf Bifurcation, where the SA phase appears to be generally replaced by an extended BS phase~\cite{szczesny_movies,szczesny_PhD,SMR_axv14}. Yet, it has also been found that when $p\gg z$ an Eckhaus-like far-field break-up of the spiral waves occurs away from the Hopf bifurcation, very much like in Fig.~\ref{mobandcom}~(bottom), see Ref.~\cite{SMR_axv14}.

\section{Evolutionary games with spontaneously emerging cyclic dominance}
\label{other_rps}

Cyclic dominance occurs not only in RPS and related evolutionary games where the closed loop of dominance is explicitly imposed on the competing strategies through the directions of invasions between them (see Fig.~\ref{web}). Quite remarkably, cyclical interactions may occur spontaneously in any evolutionary game where the competing strategies are three or more, and it is also possible to observe them in two-strategy games if certain player-specific properties are time-dependent \cite{szolnoki_pre10b}. Well-known examples include the classic public goods game \cite{perc_jrsi13}, where in addition to cooperators and defectors the third competing strategy is volunteering \cite{hauert_s02, szabo_prl02, semmann_n03, szabo_pre04b, chen_y_pre08, song_qq_pa11}, rewarding \cite{sigmund_pnas01, szolnoki_epl10, forsyth_jmb11, sasaki_pnas12, szolnoki_njp12}, or punishing \cite{boyd_pnas03, fowler_pnas05, brandt_pnas06, helbing_ploscb10, amor_pre11, deng_k_tpb12}. Further examples include the ultimatum game with discrete strategies \cite{szolnoki_prl12}, the public goods game with pool punishment \cite{hilbe_prsb10, szolnoki_pre11, sigmund_jtb12, zhang_by_expecon14}, as well as the public goods game with correlated positive (reward) and negative (punishment) reciprocity \cite{szolnoki_prx13}.

\begin{figure}
\centerline{\epsfig{file=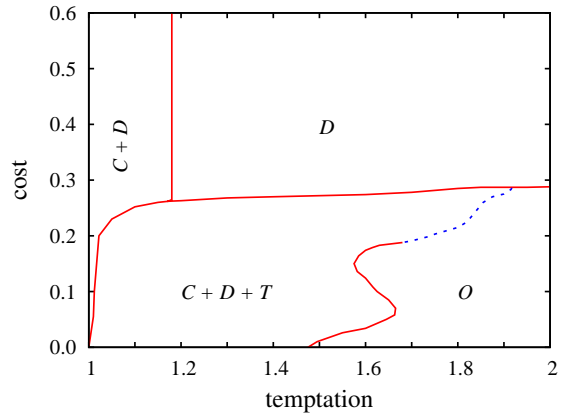,width=8.2cm}}
\caption{Phase diagram of an evolutionary prisoner's dilemma game with the ``tit-for-tat'' ($T$)
strategy, as obtained on a random regular graph. In addition to the pure defection ($D$) phase, the stationary two-strategy $C+D$ phase, and the stationary three-strategy $C+D+T$ phase, there also exists a parameter region where all three competing strategies coexist in an oscillatory phase ($O$). The spontaneous emergence of cyclic dominance therefore sustain cooperation ($C$) even at prohibitively high temptations to defect. For further details we refer to \cite{szolnoki_pre09b}.}
\label{oscillation_b}
\end{figure}

As in the original RPS game, in these evolutionary games too the spontaneous emergence of cyclic dominance promotes the coexistence of competing strategies, and through that it stabilizes solutions that would otherwise be unstable. As such, the spontaneous emergence of cyclic dominance is one of the main driving forces behind complex pattern formation, which in turn is responsible for the many differences between evolutionary outcomes reported in well-mixed and structured populations. Examples include the stabilization of reward \cite{szolnoki_epl10} and punishment \cite{helbing_ploscb10} in structured populations, the evolution of empathy and fairness in human bargaining \cite{szolnoki_prl12}, as well as an oscillatory coexistence of defection and cooperation with the ``tit-for-tat'' ($T$) strategy \cite{nowak_n92a, imhof_pnas05, szolnoki_pre09b}. The latter example is particularly instructive, as there is a transition from a three-strategy stationary state to a three-strategy oscillatory state that occurs in dependence on the cost of the ``tit-for-tat'' strategy. This phase transition is possible only in a structured population, and only because there is a spontaneous emergence of cyclic dominance between the three competing strategies. The complete phase diagram is shown in Fig.~\ref{oscillation_b}, where the dotted blue line delineates to two three-strategy phases. It is worth emphasizing that an oscillatory state emerges not because of a time-varying interaction network (the importance of which we have emphasized in Subsection~\ref{topology} in the realm of the RPS
game), but solely due to the relations between the competing strategies as defined via the payoff matrix.

Another example that merits attention is the stabilization of punishment in the realm of the public goods game. In general, punishers bear an additional cost if they wish to sanction defectors, and thus face evolutionary pressure not only from defectors, but also from pure cooperators, which effectively become second-order free-riders \cite{panchanathan_n04, fehr_n04, fowler_n05b}. Accordingly, in well-mixed populations punishment is evolutionary unstable in the absence of additional mechanisms or further strategic complexity. In structured populations, however, punishment alone introduces cyclic dominance such that punishing cooperators are able to invade defectors, defectors are able to invade pure cooperators, and pure cooperators are able to invade punishers \cite{helbing_ploscb10}. Even more complex evolutionary outcomes are possible in case of pool punishment \cite{szolnoki_pre11}, where one ``strategy'' in the closed loop of dominance can be an alliance of two strategies (for further details we refer to Subsection~\ref{three+}). Interestingly, the introduction of self-organized punishment, when the fine increases proportional with the activity of defectors, prevents the spontaneous emergence of cyclic dominance and results in a more effective way of punishment \cite{perc_njp12}. In the latter case, however, the stability of punishers is not directly challenged by pure cooperators (second-order free-riders), as the punishing activity (and the additional cost of sanctioning) seizes in the absence of defectors.

\subsection{Time-dependent learning}
\label{time}

\begin{figure}
\centerline{\epsfig{file=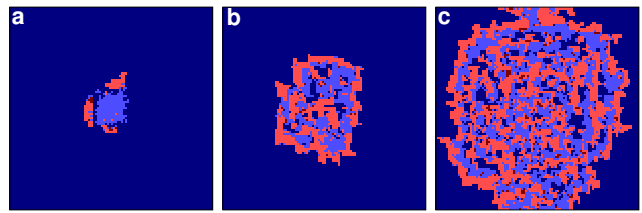,width=8.5cm}}
\caption{Spontaneous emergence of cyclic dominance due to the introduction of time-dependent learning ability of players in the spatial prisoner's dilemma game. Cooperators (blue) and defectors (red) become entailed in a close loop of dominance, which is mediated by the time-varying learning ability. Pale (dark) shades of blue and red encode low (high) learning ability of cooperators and defectors, respectively. Traveling waves emerge from an initial cluster of cooperators located in the center of the square lattice (not shown). From there on, the waves spread across the whole population [from panel (a) towards (c)], thus allowing the survival of cooperators even at very high temptations to defect. For further details we refer to \cite{szolnoki_pre10b}.}
\label{dynamic}
\end{figure}

As we have noted above, cyclical interactions may occur spontaneously not only if the competing strategies are three or more, but also in two-strategy games if certain player-specific properties are time-dependent. One such example concerns time-dependent strategy
learning ability of players in the spatial prisoner's dilemma game \cite{szolnoki_pre10b}. In a preceding work \cite{szolnoki_pre09}, it has been shown that a limited teaching activity of players after a successful strategy pass modifies the propagation of strategies in a biased way and can promote the evolution of cooperation in social dilemmas. A significantly different effect, however, can be observed if the learning ability of players is time-dependent \cite{szolnoki_pre10b}. A temporarily impaired learning activity generates cyclic dominance between defectors and cooperators, which helps to maintain the diversity of strategies via propagating waves, as illustrated in Fig.~\ref{dynamic}. Consequently, cooperators can coexist with defectors even at extremely adverse conditions. This result is particularly inspiring because it indicates the possibility of a time-dependent player-specific property to effective replace a third strategy in a closed loop of dominance. Hence, cyclic dominance can emerge between two competing strategies only. Other works on dynamical learning in games with cyclic interactions, where agents sample a finite number of moves of their opponents between each update, have shown how demographic fluctuations can lead to noise-sustained cyclic orbits~\cite{galla_prl09, realpe_jstat12}.

\subsection{Voluntary participation}
\label{volunteer}

Evolutionary games with voluntary participation where among the first to record the spontaneous emergence of cyclic dominance outside the realm of the traditional RPS formalism. In particular, the coexistence of cooperators and defectors due to the presence of volunteers (or loners) was initially reported in well-mixed populations \cite{hauert_s02, semmann_n03}, where the cyclic invasions among the three competing strategies resulted in an oscillatory state. Inspired by Alice's Adventures in Wonderland, where the Red Queen found that it takes a lot of running to stand still, the emergence of oscillations that were needed to sustain cooperation was dubbed accordingly as the ``Red Queen'' effect. Notably, the terms has been proposed in the early 70's \cite{van-valen_et73} as an evolutionary hypothesis based on which organisms must constantly adapt and evolve, not merely to gain a reproductive advantage, but also simply to survive.

Today, volunteering is known as a viable mechanisms to avert the tragedy of the commons \cite{hardin_g_s68} within the realm of the public goods game. Hauert
et al.~\cite{hauert_s02} studied this phenomenon by assuming that individuals have three available strategies, namely unconditional cooperation, defection and volunteering, which resulted in a RPS-like evolutionary dynamics. In particular, if there are $n_{c}$ cooperators, $n_{d}$ defectors, and $n_{l}$ loners in the group, then actually only $k=n_{c}+n_{d}$ individuals are involved in the public goods game. Defectors thus get the payoff $rn_{c}/k$, where $r>1$ is the multiplication factor that takes into account the synergetic effects of a group effort, while cooperators gain $(rn_{c}/k)-1$ (here $-1$ is due to their contribution to the common pool, which defectors withhold). At the same time, loners who do not participate in the game, get a fixed small payoff $\sigma$, such that $0 < \sigma < r-1$. Based on numerical simulations and the analysis of the replicator equation, it quickly follows that frequencies of all three strategies exhibit oscillations over time if $r>2$. These theoretical results have been confirmed by human experiments performed by Semmann et al.~\cite{semmann_n03}, which revealed that volunteering reduces the size of groups engaged in the public goods game, and that cooperation is indeed promoted through time-dependent oscillations. The rationale is that defectors outperform cooperators in a large predominantly cooperative group, yet as soon as defectors become the predominant force, it pays more to be a loner. But as loners grow in numbers and the actual size of the group participating in the game decreases, cooperation again becomes viable, and so the loop of dominance closes.

\begin{figure}
\centerline{\epsfig{file=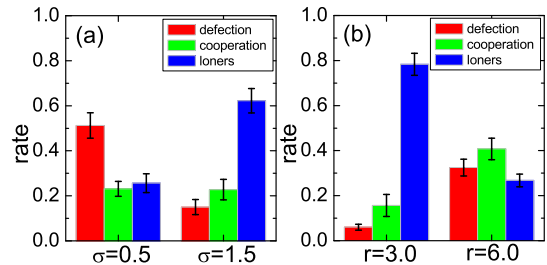,width=8cm}}
\caption{Widespread coexistence of cooperators, loners and defectors in an experimental public goods game with volunteering. Columns displays the time-averaged frequencies that were chosen by  the $90$ participants of this human experiment, depending on the payoff of loners $\sigma$ and the multiplication factor $r$. (a) The multiplication factor $r=3.0$ was set fixed, while two different values of $\sigma$ were considered (see figure legend). (b) The payoff of loners $\sigma=2.5$ was set fixed, while two different values of $r$ were considered (see figure legend). Importantly, the coexistence of all three strategies is significantly more widespread than predicted by theory, which may be attributed to the bounded rationality and emotions that are not accounted for in simulations. The whiskers in both panels show the standard error of the displayed time-averaged frequencies (rates).}
\label{social_RPS}
\end{figure}

Motivated by the pioneering human experiments \cite{semmann_n03}, a new experiment has been designed and conducted with a total of $90$ undergraduate students from Wenzhou University, who participated in repeated public goods games with volunteering in groups of five. The principal goal was to measure the average frequencies of strategies over time. The payoffs were the same as described in the preceding paragraph, thus designating $\sigma$ and the multiplication factor $r$ as the two main parameters. The results are summarized in Fig.~\ref{social_RPS}, which confirm that all three strategies coexist within a wide range of parameter values. However, the time-averaged frequencies depend on the actual values of $\sigma$ and $r$. As shown in both panels of Fig.~\ref{social_RPS}, defectors are most common for $\sigma=0.5$ and $r=3.0$, which reverses in favor of loners for $\sigma=1.5$ and $r=3.0$, thus experimentally confirming the expected promotion of loners as $\sigma$ increases. The dominance of loners is even more remarkable for $\sigma=2.5$ and $r=3.0$, but shifts towards cooperators for $\sigma=2.5$ and $r=6.0$, as shown in Fig.~\ref{social_RPS}(b). This in turn confirms the evolutionary advantage of cooperators that is warranted by higher multiplication factors. Interestingly, the outcome of these experiments indicate that the coexistence of the three strategies is more widespread than predicted by theory. For example, all three strategies coexist even for $\sigma > r-1$ [see $\sigma=2.5$ and $r=3.0$ case in Fig.~\ref{social_RPS}(b)], which may be attributed to the bounded rationality and emotions of participants that are not accounted for in simulations.

Aside from the theoretical explorations of the public goods game with volunteering in well-mixed populations and the above-described human experiments, the public goods game with volunteering in structured populations received its fare share of attention as well. In the light of the reviewed results on the spatial RPS game, it is perhaps little surprising that structured populations also maintain the survival of all three competing strategies, but they do so not through the emergence of oscillations, but rather through time-wise constant fractions of cooperators, defectors and loners \cite{szabo_prl02}. However, the topology of the interaction network can play a key role, much the same as by the RPS game \cite{szabo_jpa04, szolnoki_pre04b}. Specifically, if the interaction network has sufficiently strongly pronounced small-world properties (there exists a threshold in the fraction of shortcut links that must be exceeded), then the stationary three-strategy phase gives way to an oscillatory three-strategy phase \cite{szabo_pre04b, wu_zx_pre05}. Further strengthening the analogies between the RPS game and the public goods game with volunteering, we note that  oscillations in the latter can also be evoked not only through changes in the topology of the interaction network, but also through changes in the payoffs, as illustrated in Fig.~1 of \cite{wu_zx_pre05}.

Interestingly, the so-called joker strategy, introduced by Arenas et al.~\cite{arenas_jtb11}, is conceptually different from the loner strategy, but has a similar impact on the evolution of cooperation in social dilemmas. In particular, in finite well-mixed populations oscillations emerge spontaneously, and this irrespective of the system size and the type of strategy updating \cite{arenas_jtb11,requejo_pre12b}, thus adding yet further support to the relevance of cyclic dominance in evolutionary games.

\subsection{When three competing strategies are more than three}
\label{three+}

If the competing strategies in an evolutionary game are more than two (with some exceptions, as described in Subsection~\ref{time}), this opens up the possibility for the spontaneous emergence of cyclic dominance with much the same properties that characterize the classical RPS
game. Sometimes, however, three competing strategies are more than three, because some subsystem solutions, typically composed of a subset of the three original strategies, manifest as an additional ``strategy''. The fourth ``strategy'' is not exactly equivalent to the original three strategies because it only emerges during the evolutionary process, yet its involvement in the closed loop of dominance is just as important. These rather exotic solutions, which can be observed only in structured populations, have recently been reported in the spatial public goods game with pool punishment \cite{szolnoki_pre11}, and in the spatial ultimatum game with discrete strategies \cite{szolnoki_prl12}.

\begin{figure}
\centerline{\epsfig{file=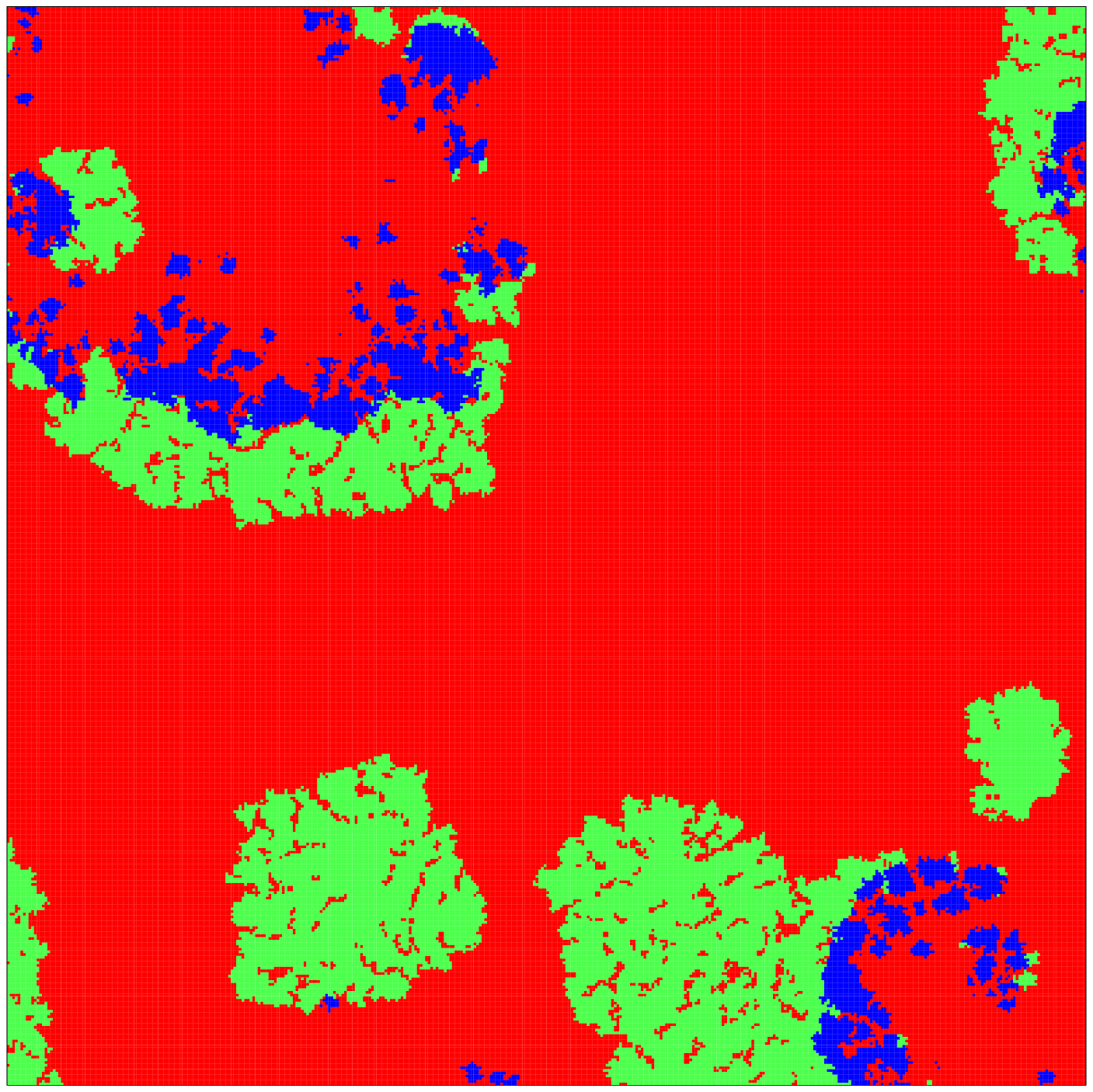,width=4cm} \epsfig{file=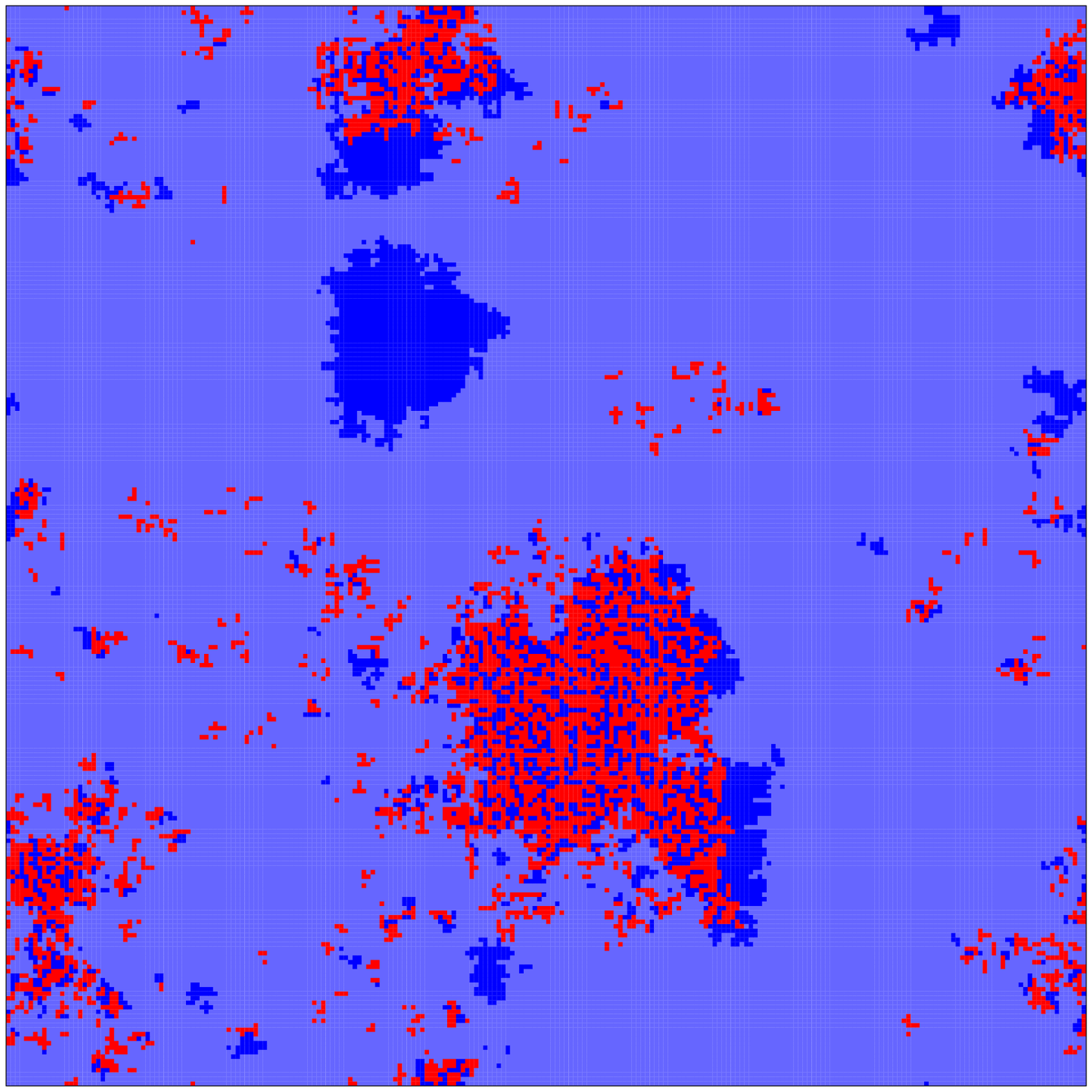,width=4cm}}
\caption{In structured populations three competing strategies can in fact be more than three. The spontaneous emergence of cyclic dominance can be driven by pure strategies, as well as by stable alliances of a subset of pure strategies, which in the closed loop of dominance act just the same. Left panel: Cyclic dominance between cooperators (blue), defectors (red), and a stable alliance between defectors and pool punishers (red and green). Here defectors beat cooperators, who beat the alliance of defectors and pool punishers, who in turn beat pure defectors \cite{szolnoki_pre11}. Note that one ``strategy'' in the closed loop of dominance is thus an alliance of two strategies. Right panel: Cyclic dominance in the spatial ultimatum game with discrete strategies \cite{szolnoki_prl12}. Two emphatic strategies, $E_1$ (light blue) and $E_2$ (navy blue), compete with a strategy $A$ (red) that is characterized by a particular $(p,q)$ pair. Here $E_1$ beats the stable alliance of $E_2+A$, who in turn beats $E_2$, who in turn beats $E_1$,thus closing the loop of dominance. As in the spatial public goods game with pool punishment, here to one ``strategy'' in the loop is an alliance of two strategies.}
\label{pool}
\end{figure}

In the spatial public goods game with pool punishment, we initially have three competing strategies, which are cooperators ($C$), defectors ($D$), and pool punishers ($P$). Quite remarkably, at certain parameter values (see \cite{szolnoki_pre11} for details), defectors and pool punishers form a stable alliance, which effectively manifests as a strategy, and this strategy forms a closed loop of dominance with cooperators and defectors. Cyclic dominance ensues, where in addition to the two pure strategies $C$ and $D$, the $D+P$ alliance is the third competing strategy. As a consequence, the characteristic spatial patterns feature propagating fronts made up of all the mentioned competitors, as illustrated in the left panel of Fig.~\ref{pool}.

The peculiarity of the above-described solution indicates that it may not be satisfactory to describe the actual state of a three-strategy system by a single ``point'' in a ternary phase diagram. Namely, just the fractions of the competing strategies do not necessarily determine the actual state of the system in an unambiguous way, and this is because spatiality offers an additional degree of freedom in comparison to solutions that can be observed in well-mixed populations. As described in \cite{szolnoki_pre11}, this new degree of freedom may give rise to an additional ``strategy'' that is as such not present at the start of the evolutionary process.

Importantly, although such evolutionary solutions may at first glance appear as special and perhaps even limited outcomes of a particular evolutionary game, we emphasize that they may in fact be common place. To corroborate this, we review a completely different evolutionary game, namely the spatial ultimatum game with discrete strategies \cite{szolnoki_prl12}. Unlike in the public goods game, in the ultimatum game the focus is not on the aversion of the tragedy of the commons, but rather on the evolution of empathy and fairness \cite{sigmund_sa02}. While discrete strategies still cover the whole $(p,q)$ proposal-acceptance parameter plane, the discreetness takes into consideration the fact that in reality these levels are coarse-grained rather than continuous. Thus, instead of the infinite number of continuous strategies, we have $N$ discrete strategies $E_i$, defined according to their $(p,q)$ interval (see \cite{szolnoki_prl12} for details). If $p=q$, we have $N$ different classes of empathetic players who can compete with an additional strategy $A$ that is characterized by an arbitrary $(p,q)$ pair. Most crucially, the finite number of competing strategies enables pattern formation, and as illustrated in the right panel of Fig.~\ref{pool}, cyclic dominance may emerge spontaneously where a member in the closed loop of dominance can be an alliance of two strategies. This example, in addition to the previously described one concerning the spatial public goods game with pool punishment \cite{szolnoki_pre11}, thus corroborates the fact that in structured populations three competing strategies can
in fact be more than three.

\section{Cyclic dominance between more than three strategies}
\label{four}

Extensions of the classical RPS game to more than three competing strategies are straightforward (see Fig.~\ref{web}), and in addition to their obvious appeal from the theoretical point of view \cite{frachebourg_jpa98, szabo_pre01b, szabo_jpa05, rulquin_pre14, mowlaei_jpa14}, such games also find actual applicability in describing competition in microbial populations \cite{szabo_pre01a} and larger ecological food webs and communities \cite{berlow_jae04, allesina_s08, allesina_te08, allesina_pnas11, stouffer_s12, knebel_prl13}. Based on the review of results concerning the RPS and related three-strategy evolutionary games with cyclic dominance, it is little surprising that the complexity of spatial patterns, and indeed the complexity of evolutionary solutions in general, increases drastically as we increase the number of strategies that form a closed loop of dominance. The spontaneous emergence of defensive alliances and numerous stable spatial distributions that are separated by first and second-order phase transitions are just some of the phenomena that one is likely to observe.

The reason behind the mushrooming number of stable solutions with the rising number of competing strategies lies in the fact that the solutions of subsystems, where some strategies are missing, can also be solutions of the whole system. The final stationary state can thus be determined not only by the competition of individual strategies, but also by the competition of subsystem solutions (defensive alliances) that are characterized by their own composition and spatiotemporal distribution of strategies (we have already reviewed two such examples in Fig.~\ref{pool}). Consequently, the general understanding of evolutionary games with many competing strategies requires the systematic analysis and comparison of all possible subsystem solutions, i.e., solutions, that are formed by just a subset of all available strategies.

\begin{figure}
\centerline{\epsfig{file=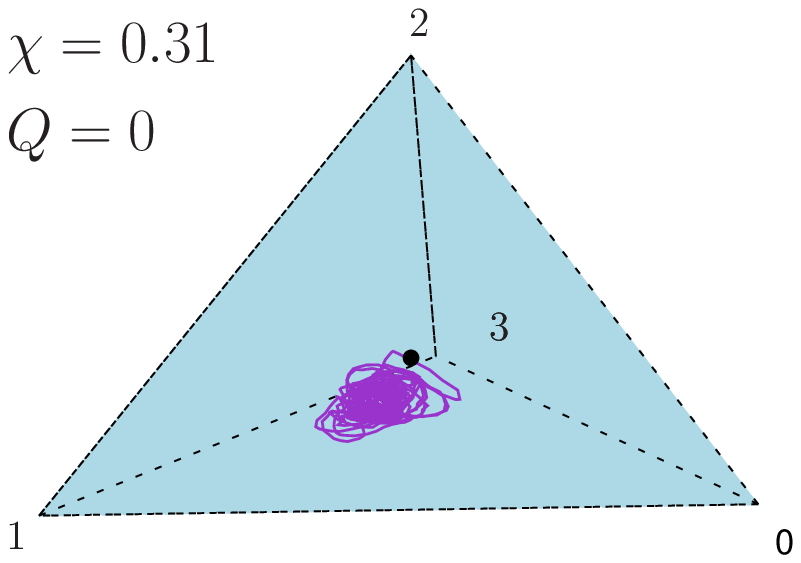,width=2.83cm}\epsfig{file=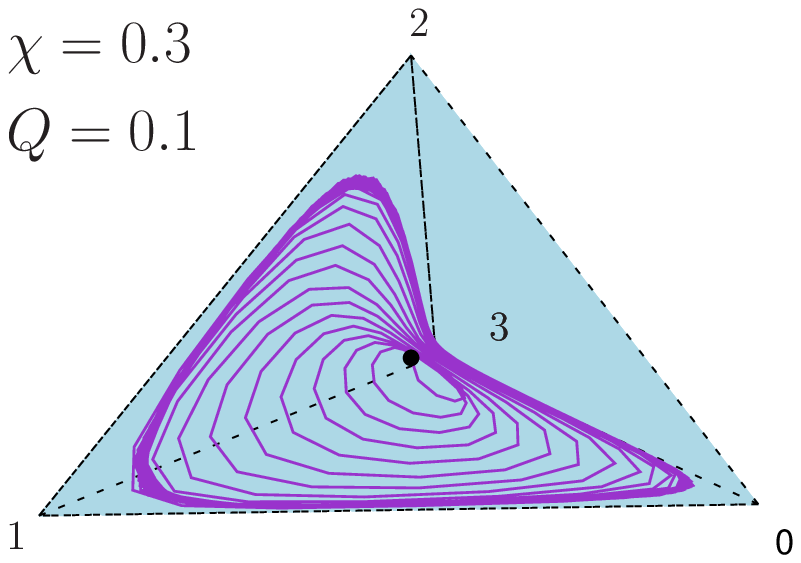,width=2.83cm}\epsfig{file=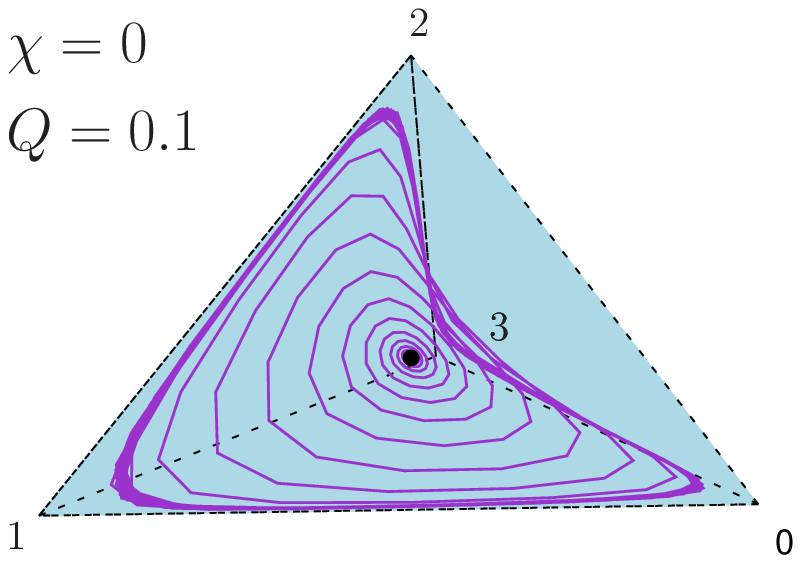,width=2.83cm}}
\caption{Globally synchronized oscillations in a four-strategy cyclic dominance game \cite{rulquin_pre14}. Depicted are simplex representations of the time evolution of the four densities for three different combinations of the fraction of long-range links
$Q$ and the heterogeneity of the food web $\chi$ (see legend). Starting from the initially random homogenous state (black dot), the system evolves (left) towards finite size, stochastic fluctuations around the asymptotically stable fixed point (the amplitude of these fluctuations decreases with increasing system size \cite{lutz_jtb13}), (middle) towards global oscillations, where the asymptotic state is a limit cycle whose perimeter may be used as an order parameter for the transition, or (right) towards global oscillations, where the orbit approaches the heteroclinic one and the average perimeter is close to the maximum value. We refer to \cite{rulquin_pre14} for further details.}
\label{global}
\end{figure}

The simplest generalization of the RPS game is towards a four-strategy cyclic dominance game \cite{peltomaki_pre08, case_epl10, durney_pre11, intoy_jsm13}. As reported in \cite{peltomaki_pre08}, the formation of spiral patterns takes place only without a conservation law for the total density, and like in the three-strategy case, strong mobility can destroy species coexistence. Cyclic dominance among four strategies also gives rise to so-called ``neutral'' strategies, who do not invade each other --- an option that is not available among three competing strategies. In the absence of mixing, neutral strategies can easily result in population-wide frozen states, where only those strategies are present who do not invade each other. More precisely, the coexistence of all four strategies is limited only to a rather narrow range of invasion rates, whereas otherwise the system evolves into a frozen state where only neutral strategies are present \cite{szabo_pre08}. This observation further supports the fact that the topology of the food web alone cannot unambiguously determine the evolutionary outcome of cyclical interactions. In addition, the invasion rates and the interaction range are likewise crucial parameters that determine the transitions from coexistence to uniformity among cyclically competing strategies \cite{lutz_jtb13, knebel_prl13, szolnoki_xxx14}.

Globally synchronized oscillations in a four-strategy cyclic dominance game are also possible, as reported in \cite{rulquin_pre14} and presented in Fig.~\ref{global}. In fact, Rulquin and Arenzon \cite{rulquin_pre14} have shown that such coexistence phases require less long-range interactions than three-strategy phases, although intuitively one would expect the opposite. Since the emergence of long distance, global synchronization states is thus possible even when the food web has multiple subloops and several possible local cycles, this leads to the conjecture that global oscillations might be a general characteristic, even for large, realistic food webs.

In addition to the coexistence of all competing strategies, the emergence of ``frozen'' states, and global oscillations mentioned above, curvature driven domain growth is also possible when the evolutionary dynamics is governed by interactions of equal strength among more than three strategies \cite{avelino_pre12, avelino_pre12b, roman_jsm12}. In this case, the time dependence of the areas of different domains, which can be quantified effectively by means of the length of domain walls, conforms to the well-known algebraic law of the domain growth process \cite{bray_ap94}.

\subsection{Alliances}
\label{alliances}

The subject that deserves particular attention within the realm of cyclic dominance between more than three strategies is the formation of alliances. As described in the introductory paragraphs of this section, alliances play a key role in the fascinating complexity and multitude of solutions that grace such systems, particularly through the formation of subsystems solutions that together with the pure strategies form the complete solutions of the whole system.

Alliances can emerge in strikingly different ways, although the most common one is the formation of so-called defensive alliances. For example, in a four-strategy predator-prey system, two neutral strategies can defend each other from the two external invaders \cite{szabo_pre04}, hence the name a ``defensive alliance''. The protective role of an alliance can also be observed when three cyclically dominating strategies are present. In this case, they can protect each other from an external invader by invading the predator of their prey \cite{szabo_pre01b}. We also emphasize that cyclically dominant strategies may not necessarily form defensive alliances, even if the governing food web is complex and large. For example, in Fig.~\ref{web}(d) the strategies $RS+KS+SS$ form a subsystem solution, but this is not an alliance as it is vulnerable against the invasion of an external strategy. On the other hand, the strategies $RS+KR+SK$ form an actual alliance, and a membership in the latter can effectively protect all three strategies within the loop from an external invasion.

Moreover, if the number of competing strategies is high enough, it is also possible to observe the competition between the different alliances \cite{szabo_jpa05}. Alliance-specific invasion rates, however, can break the symmetry of the food web, and subsequently only a single alliance survives \cite{perc_pre07b}. The complexity of a six-species
predator-prey food web with alliance-specific invasions rates also enables noise-guided evolution within cyclical interactions \cite{perc_njp07b}, which is deeply rooted in short-range spatial correlations and is conceptually related to the coherence resonance phenomenon in dynamical systems via the mechanism of threshold duality \cite{pikovsky_prl97, gammaitoni_rmp98, perc_pre05, perc_njp06a}.

\begin{figure}
\centerline{\epsfig{file=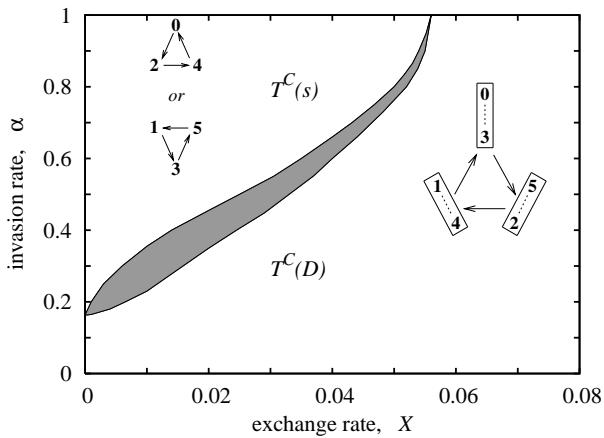,width=8cm}}
\caption{Who can form an alliance? The displayed phase diagram of stable solutions in dependence on the invasion rate ($\alpha$) and the strength of mixing ($X$)
in a six-species predator-prey system reveals that in fact all possible alliances between two or more competing strategies can be viable at specific parameter regions
\cite{szabo_pre08b}. In the upper left corner, the three-strategy cyclic alliance $0 + 2 + 4$ or the three-strategy cyclic alliance $1 + 3 + 5$ will dominate the system. In the parameter region marked by $T^C (D)$, on the other hand, three alliances, each consisting of two-neutral strategies, play the RPS game at a higher level, i.e., at the level of competing alliances (rather than pure strategies). Within the shadowed parameter region all the mentioned alliances coexist.}
\label{2_or_3}
\end{figure}

\begin{figure}
\centerline{\epsfig{file=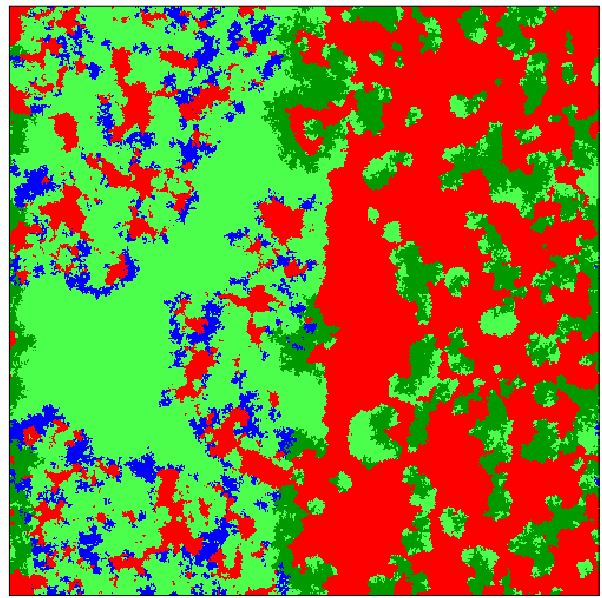,width=8cm}}
\caption{Alliances may also form spontaneously, without support from specific relations within the governing food web. This is particularly common in games that are played in groups \cite{perc_jrsi13}, where the payoffs stemming from the multi-point interactions may give rise to relations between the competing strategies that make the formation of an alliance evolutionary most advantageous. One particular such example concerns the spatial public goods game with pool ($O$) (light green) and peer ($E$) (dark green) punishment, where the three-strategy cyclic alliance $D+C+O$ and the three-strategy cyclic alliance $E+O+D$ emerge spontaneously. Subsequently, these two alliances compete for space effectively as ``strategies'', thereby giving rise to complex yet beautiful spatial patterns. Cooperators are blue while defectors are denoted red.}
\label{dco}
\end{figure}

When it comes to the study of alliances within games of cyclic dominance, the general question to answer is which strategies are capable to form alliances? As we have reviewed above, two strategies are sufficient to form an alliance, but of course three or even more strategies could also form one. The phase diagram presented in Fig.~\ref{2_or_3} demonstrates that in fact all the mentioned alliances can be present in the stable solution of the system, depending also on the parameter values that determine invasion and mixing in the studied six-species predator-prey system \cite{szabo_pre08b}. The key property that determines wether an alliance is viable or not is its ability to invade other alliances or resist such an invasion itself. Like by the competition among pure strategies, here too the appropriate approach is to measure directly the velocity of invasion between competing domains that contain specific subsystem solutions \cite{vukov_pre13}. The use of special initial conditions, like the one illustrated in Fig.~\ref{dco} (remnants of the straight vertical interface that separated two three-strategy cyclic alliances can still be clearly inferred), is thereby very useful and recommended. Unlike by pure strategies, however, here we first need to partition the population and wait for the appropriate subsystem solutions to emerge before removing the barriers along the interfaces.

Importantly, the formation of alliances is not always the consequence of specific relations within the governing food web. Alliances may also emerge spontaneously during the evolutionary process, for example due to payoff relations between the competing strategies that are due to multi-point (group) interactions. The phenomenon was observed in the spatial public goods game, where in addition to cooperators and defectors both pool ($O$) and peer ($E$) punishers competed for space \cite{szolnoki_pre11b}. Figure~\ref{dco} illustrates an example of the spontaneous emergence of two three-strategy cyclic alliances. Notably, either the $D+C+O$ or the $E+O+D$ alliance can be stable solutions of the whole system at specific values of the punishment cost and fine. But if they meet, as displayed in Fig.~\ref{dco}, the $E+O+D$ alliance will invade the $D+C+O$ alliance. Remarkably, the invasion process between the two alliances is effectively the same as it would be between two pure strategies.

As by the competition among pure strategies \cite{reichenbach_n07}, by the competition of alliances too the role of mobility (or diffusion) is likewise important. Within the framework of a nine-species predator-prey model \cite{szabo_pre01b}, it has been shown that below a critical mobility rate, the system exhibits expanding domains of three different defensive alliances, each consisting of three cyclically dominant species. Due to the neutral relationship between these three defensive alliances, a voter-model type coarsening starts \cite{dornic_prl01}, and after a sufficiently long time only one alliance survives if the system has finite size. Above the critical mobility rate, however, the three defensive alliances can all coexist \cite{szabo_jtb07}. In a way, the effect of high mobility is thus the opposite to the one observed in the RPS game. There high mobility rates jeopardize diversity and lead to the survival of a single strategy \cite{reichenbach_n07}, while the coexistence of defensive alliances may receive unexpected support. Further related to the impact of mobility on the survival of alliances, a recent study has demonstrated that mobility-dependent selection can prefer either a three-strategy cyclic alliance or a neutral pair of two-strategy alliances within a four-strategy system \cite{dobrinevski_pre14}.

At this point, we again emphasize that it is not enough to simply separate random mixtures of selected strategies that are involved in particular alliances and monitor the evolution. Before the barriers along the interfaces are removed, a sufficiently long relaxation time has to be taken into account, which allows the actual subsystem solutions (rather than just the random mixtures of the involved strategies) to emerge. Only then will the actual competition between the alliances unfold and give correct insights into the power relations between them. In the absence of such a procedure, it may easily happen that one of the considered subsystem solution cannot evolve fast enough, for example due to requiring a longer relaxation time than another subsystem solution, and thus it becomes impossible to properly compare their viability. A specific example to that effect concerns a five-strategy predator-prey system \cite{kang_pa13, vukov_pre13}, where if the mobility is strong enough the dominance between the competing alliances can vanish simultaneously. This phenomenon is then also accompanied by a divergence in the density fluctuations of individual strategies. In the absence of mobility, however, the dominance between the alliances does not vanish simultaneously, and only a sharp peak in the density fluctuations can be observed instead of the power-law divergence. The phenomena observed in the realm of this five-strategy predator-prey system in fact outline a more general type of behavior, namely the mobility-induced reversal of the direction of invasion between the competing alliances, thus highlighting yet again that the topology of the food web does not always (in fact it does so only rarely) determine the final evolutionary outcome in structured populations.

\section{Conclusions and outlook}
\label{final}

As we hope this review clearly shows, cyclical interactions are a fascinating subject which has the power to captivate with the complexity and beauty of the governing evolutionary dynamics, and which in addition to this has many real-life applications that we have only begun to fully appreciate. We have reviewed main results in well-mixed populations with the focus on oscillatory dynamics and the extinction to absorbing states due to finite system size.
We have also reviewed results on the classical RPS game in structured populations. We have first focused on the important role of the topology of the interaction networks, which may induce oscillatory states, although the latter can be evoked by changes in the game parametrization too. We have also focused extensively on mobility, which has many practical implications as in real life prey and predators are frequently on the move in order to maximize their chances of survival. We have emphasized that mobility can either promote or jeopardize biodiversity, and it may give rise to fascinatingly rich pattern formation, including spiral and target waves, multi-armed spirals with two, three and four arms, as well as anti-spirals. Mobility can also affect the basins of coexistence and extinction through intra-species and inter-species epidemic spreading. An alternative metapopulation modeling approach is also reviewed, where the distinction of pair-exchange and hopping processes allows us to discriminate movement in crowded and in diluted regions. The resulting nonlinear mobility promotes the far field breakup of spiral waves and enhances their convective instability. Within this metapopulation formulation, we have also derived perturbatively the complex Ginzburg-Landau equation.

In addition to the review of the well-mixed and spatial rock-paper-scissors game,
we have also given special attention to evolutionary games where cyclic dominance emerges spontaneously, for example between only two competing strategies through time-dependent learning, among three strategies in public goods game with volunteering, punishment or reward, as well as by means of the introduction of the joker strategy.
We have reviewed
how alliances may effectively act as strategies and close the loop of dominance in three-strategy evolutionary games, including the
public goods game with pool punishment and the
ultimatum game with discrete strategies. From there onwards, we have surveyed cyclic dominance when the competing strategies are more than three,
where we have emphasized the possibility of
competition between alliances. We have highlighted the importance of the correct preparation of special initial conditions that allow proper monitoring of the invasion relations between the alliances, and we have given prominent exposure to the fact that alliances
may also form spontaneously, without support from specific relations within the governing food web.

There are still unexplored problems related to cyclic dominance in evolutionary games that merit further attention. While physics-inspired studies account for the majority of recent theoretical advances on this topic, there also exist much experimental and theoretical work that was outside the scope of this review. Competitive intransitivity and the coexistence of species in intransitive networks \cite{laird_an06, laird_e08}, as well as community-scale biodiversity in ecological systems with a large number of species \cite{kendall_e99} and the governing large competitive networks \cite{allesina_s08, allesina_te08, allesina_pnas11}, are just some of the topics that we did not cover in detail. While we have mentioned several experiments and empirical data that corroborate the importance of cyclical interaction in the Introduction, covering this in detail could easily amount to an independent review. In addition to the often-cited examples mentioned in the Introduction, most recent experimental research suggests that we may have only just begun to scratch the surface of the actual importance of cyclic dominance in nature \cite{kaspari_s14, lebrun_s14}.
It would be desirable for the fundamental theoretical advances to become more in sync with the experiments by means of more dedicated and precise mathematical modeling.
It is worth noting
that cyclic dominance could be crucial not just to understand biodiversity, but also social diversity \cite{santos_n08, perc_pre08, santos_jtb12}.

In terms of future fundamental research, we would like to point towards the basically unexplored subject of cyclic dominance in group interactions. In social dilemmas, the transition from pairwise to group interactions \cite{pacheco_prsb09, santos_md_jtb12} brings with it many qualitative changes in the evolutionary dynamics (see \cite{perc_jrsi13} for a review), and the reasonable expectation is that the same would hold true in games of cyclic dominance too. We have recently considered the impact of different interaction ranges in the spatial RPS
game, and found that the transition from pairwise to group interactions can decelerate and even revert the direction of the invasion between the competing strategies \cite{szolnoki_xxx14}. These results thus indicate that, in addition to the invasion rates and other properties of cyclic dominance games already reviewed here, the interaction range is at least as important for the maintenance of biodiversity among cyclically competing strategies, and thus deserves further attention.

Motivation for future research can also be gathered from coevolutionary games \cite{perc_bs10}, where cyclical interactions can still be considered as being at an early stage of development. While initially many studies that were performed within the realm of social dilemma games appeared to be valid also for games that are governed by cyclical interactions, recent research has made it clear that this is by far not the case. The incentives are thus there to reexamine the key findings, which were so far reported only for pairwise games on coevolutionary networks, also for games that are governed by cyclical interactions. We conclude by noting that devising and studying cyclic dominance systems with greater practical applicability and predictive power, including more complex interaction networks and larger numbers of competing strategies, should also be well within the scope of viable forthcoming research.

\begin{acknowledgments}
This work was supported by the Hungarian National Research Fund (Grant K-101490), the National Natural Science Foundation of China (Grants 61203145 and 11047012), and the Slovenian Research Agency (Grants J1-4055 and P5-0027). Bartosz Szczesny is thankful for the support of an EPSRC studentship (Grant No. EP/P505593/1). The authors gratefully acknowledge the fruitful contributions of many collaborators to part of the research reviewed here.
\end{acknowledgments}

\end{document}